# Token-Mol 1.0: tokenized drug design with large language model

Jike Wang[1,#], Rui Qin[1,#], Mingyang Wang[1,#], Meijing Fang[1], Yangyang Zhang[1], Yuchen Zhu[1], Qun Su[1], Xiaozhe Wan[2], Qiaolin Gou[1], Chao Shen[1], Odin Zhang[3], Zhenxing Wu[1], Dejun Jiang[1], Xujun Zhang[1], Huifeng Zhao[1], Jingxuan Ge[1], Zhourui Wu[4], Liwei Liu[2,*], Yu Kang[1,*], Chang-Yu Hsieh[1,*], Tingjun Hou[1,*]

[1]*College of Pharmaceutical Sciences, Zhejiang University, Hangzhou 310058, Zhejiang, China*

[2]*Advanced Computing and Storage Laboratory, Central Research Institute, 2012 Laboratories. Huawei Technologies Co., Ltd., Nanjing 210000, Jiangsu, China*

[3]*Paul Allen School, University of Washington, Seattle*

[4]*Key Laboratory of Spine and Spinal cord Injury Repair and Regeneration, Ministry of Education, Tongji University, Shanghai, China;*

[#]*Equivalent authors*

## Corresponding authors

**Tingjun Hou**

E-mail: tingjunhou@zju.edu.cn

**Chang-Yu Hsieh**

E-mail: kimhsieh@zju.edu.cn

**Yu Kang**

E-mail: yukang@zju.edu.cn

**Liwei Liu**

E-mail: liuliwei5@huawei.com

# Abstract

Significant interests have recently risen in leveraging sequence-based large language models (LLMs) for drug design. However, most current applications of LLMs in drug discovery lack the ability to comprehend 3D structures, thereby limiting their effectiveness in tasks that specifically involve molecular conformations. In this study, we introduce Token-Mol, a token-only 3D drug design model. This model encodes all pertinent molecular information, including both 2D and 3D structures, as well as molecular property data, into tokens. It transforms classification and regression tasks in drug discovery into probabilistic prediction problems, ultimately enabling a comprehensive and cohesive learning approach through a unified paradigm. Token-Mol is constructed on the foundation of the transformer decoder architecture, and trained using random causal masking techniques. Moreover, we propose the Gaussian cross-entropy loss function to overcome the challenges encountered in regression tasks, which significantly enhances the capacity of LLMs to learn continuous numerical values. Experimental validations demonstrate that Token-Mol surpasses state-of-the-art methods in the molecular conformation generation task, achieving improvements across various metrics of over 10% and 20% on two datasets, respectively. In a series of molecular property prediction tasks, Token-Mol exhibits an average improvement of 30% in regression tasks compared to existing token-only models. In the pocket-based molecular generation task, Token-Mol can generate molecules with a notable improvement in the drug-likeness (QED) and synthetic accessibility by ~11% and 14%, respectively, meanwhile maintaining a comparable level of affinity, as estimated by the Vina score, compared to existing methods. When compared to expert models based on diffusion architecture that exhibit similar performance, Token-Mol demonstrates an impressive generation speed that is 35 times faster. In real-world drug design scenarios, Token-Mol's generated candidates exhibit a noteworthy 1-fold enhancement in average success rate across tests on 8 specific targets. A multiple optimizations of binding affinity and drug-like properties are further achieved through the incorporation of reinforcement learning, with the average affinity increased by 18% and the QED scores by at least 20%. Token-Mol effectively overcomes the precision limitations of token-only models and has the potential to seamlessly integrate with more generalized models like ChatGPT. This, in turn, paves the way for the development of a universal artificial intelligence drug design model that facilitates rapid and high-quality drug design by experts.

**Introduction**

Drug discovery traverses a remarkably intricate journey. Recent years have witnessed profound advancements in artificial intelligence (AI) technologies, particularly deep learning (DL), which has been progressively permeating multiple facets of drug development. These technologies are emerging as critical catalysts for innovative drug research. However, the formidable cost associated with acquiring annotated data sets in drug discovery remains a significant impediment to the advancement in this field. Recently, the rapid evolution of unsupervised learning frameworks, epitomized by BERT[1] and GPT[2], has introduced unsupervised chemical and biological pre-training models across disciplines such as chemistry[3-12] and biology[13-16]. These models undergo large-scale unsupervised training to learn representations of small molecules or proteins, subsequently fine-tuned for specific applications. By leveraging unsupervised learning on large-scale datasets, these pre-training models effectively addresses the challenges associated with sparse labeling and suboptimal out-of-distribution generalization, resulting in significantly improved performance.[17].

Large-scale molecular pre-training models can be broadly categorized into two main groups: models based on chemical language and models utilizing molecular graphs. First, chemical language models encode molecular structures using representations such as simplified molecular input line entry system (SMILES)[18] or self-referencing embedded strings (SELFIES)[19]. They employ training methodologies akin to BERT or GPT, well-established in natural language processing (NLP). Notable examples include SMILES-BERT[20], MolGPT[21], and Chemformer[22], which exhibit architectural similarities to universal NLP models. Second, graph-based molecular pre-trained models exhibit higher versatility. They represent molecules in a graphical format, with nodes for atoms and edges for chemical bonds. Pre-training methodologies include various techniques, such as random masking of atom types, contrastive learning, and context prediction[23-25]. Unlike language-based models, graph-based molecular pre-trained models inherently incorporate geometric information, as demonstrated by methods like GEM[26] and Uni-Mol[27].

Despite their advancements, both classes of models exhibit distinct limitations. Large-scale molecular pre-training models based on the chemical language face a significant constraint in their inability to inherently process 3D structural information, which is pivotal for determining the physical, chemical, and biological properties of molecules. Consequently, these models are inadequate for downstream tasks that involve 3D structures, such as molecular conformation generation and 3D structure-based drug design. In contrast, graph-based molecular pre-trained models can effectively incorporate 3D information. However, existing approaches primarily focus on learning molecular representations for property prediction rather than molecular generation. Moreover, integrating these models with universal NLP models presents considerable challenges. As a result, a comprehensive model capable of addressing all drug design tasks remains elusive. Addressing the limitations of these two model types to develop a pre-trained model suitable for all drug design scenarios, and easily integrable with existing general-purpose large language models, is a pressing need.

The emergence of universal artificial intelligence offers new opportunities in this domain. By leveraging vast amounts of data, these models acquire expert knowledge across various fields, providing valuable assistance to practitioners[2, 28-30]. Recent studies suggest that GPT-4 demonstrates a profound understanding of key concepts in drug discovery, including therapeutic proteins and the fundamental principles governing the design of small molecule-based and other types of drugs. However, its efficacy in specific drug design tasks, such as *de novo* molecule generation, molecular structure alteration, drug-target interaction prediction, molecular property estimation, and retrosynthetic pathway prediction, requires further refinement[31]. Nevertheless, the application of a token-based approach by the above models to handle continuous spatial data is particularly noteworthy.

Building on this concept, Born et al. introduced the Regression Transformer[32], which integrates regression tasks by encoding numerical values as tokens. Nonetheless, this method overlooks the intricate 3D structural complexities of molecules. Additionally, Flam-Shepherd and Aspuru-Guzik proposed directly tokenizing 3D atomic coordinates (XYZ) to represent molecular 3D structures[33]. Concurrently, the BindGPT framework employs a similar approach to generate molecular structures and their corresponding 3D coordinates[34]. While the performance of these models still necessitates enhancement, both approaches have exhibited promising outcomes in relevant drug design tasks. These results highlight the potential of large models to grasp the semantics of numerical values and affirm the feasibility of employing token-only models to handle continuous data.

However, directly training language models on Cartesian coordinates of atoms presents unique challenges. Specifically, for larger molecules, the extensive XYZ coordinates can result in excessively long sequences, complicating the model's learning process. Furthermore, achieving invariance through random translation and rotation does not confer equivariance.

To overcome the limitations of current models, we present Token-Mol, a pioneering large-scale language model tailored for molecular pre-training. To enhance compatibility with existing general-purpose models, we employ a token-only training paradigm, recasting all regression tasks as probabilistic prediction tasks. Token-Mol is constructed with a 12-layer Transformer decoder architecture, integrating essential 2D and 3D structural information via SMILES and torsion angle tokens. The SMILES representation accurately captures the 2D structure of molecules, thereby mitigating errors associated with empirical bond predictions[33]. For representing 3D structural information, a wealth of literature on conformation generation cogently illustrates that the utilization of torsion angles, as opposed to the direct prediction of Cartesian XYZ coordinates, results in markedly improved outcomes[33, 35-43]. The torsion-based paradigm, capturing relative atomic positions, bestows the model with equivariance. Moreover, such a method necessitates merely a modest number of torsion tokens to embody a molecule's entire 3D structure, thus markedly curtailing sequence length and simplifying contextual intricacy for each instance during training.

Furthermore, we utilize a random causal masking strategy during pre-training, leveraging a combination of Poisson and random distributions to stochastically mask training data. This strategy enhances the model's fill-in-the-blank generation capability, increasing its adaptability to a wide range of downstream tasks. To address the token-only model's limited sensitivity to numerical values, we introduce a Gaussian cross-entropy (GCE) loss function, replacing the traditional cross-entropy loss. This innovative loss function assigns weights to each token during training, enabling the model to learn the relationships between numerical tokens. Additionally, Token-Mol demonstrates exceptional compatibility with other advanced modeling techniques, including fine-tuning and reinforcement learning (RL). This integrative capability facilitates the further optimization of its performance in downstream tasks, thereby enhancing its utility in various applications.

To validate the capabilities of Token-Mol, we conducted comprehensive assessments across molecular conformation generation, property prediction, and pocket-based molecular generation tasks.

In the molecular conformation generation task, Token-Mol outperforms existing state-of-the-art (SOTA) methods, achieving over 10% and 20% improvements on two datasets across precision metrics, thereby producing superior-quality conformations.

In the molecular property prediction task, Token-Mol achieves an average improvement of 30% in regression tasks compared to token-only models.

In the context of the pocket-based molecular generation task, Token-Mol not only achieves molecules with Vina scores comparable to those produced by state-of-the-art (SOTA) models, but also improves drug-likeness (QED) and synthetic accessibility (SA) by approximately 11% and 14%, respectively.

To further validate the generalization capability of Token-Mol, we conducted tests in real-world drug design scenarios. Notably, the drug-like molecules generated by Token-Mol demonstrated a 1-fold increase in average success rate across evaluations on 8 different targets.

Furthermore, the integration of RL into Token-Mol effectively enhances the performance of specific downstream tasks within more realistic scenarios, providing an advantage over large models based on geometric graph neural networks (GNN). For instance, we applied RL to refine the pocket-based molecular generation process, specifically two targets. Trough this process, the affinity of the compounds for their respective targets, as measured by the Vina Score, is 18% higher compared to the average Vina Score of high-affinity molecules for real molecules.

Finally, we demonstrated Token-Mol's seamless integration with general large language models through a simple dialogue example. Token-Mol stands as a groundbreaking token-only 3D molecular language model. The aforementioned findings offer compelling evidence of the inherent potential in this framework, presenting a novel outlook on standardizing AI models for drug design.

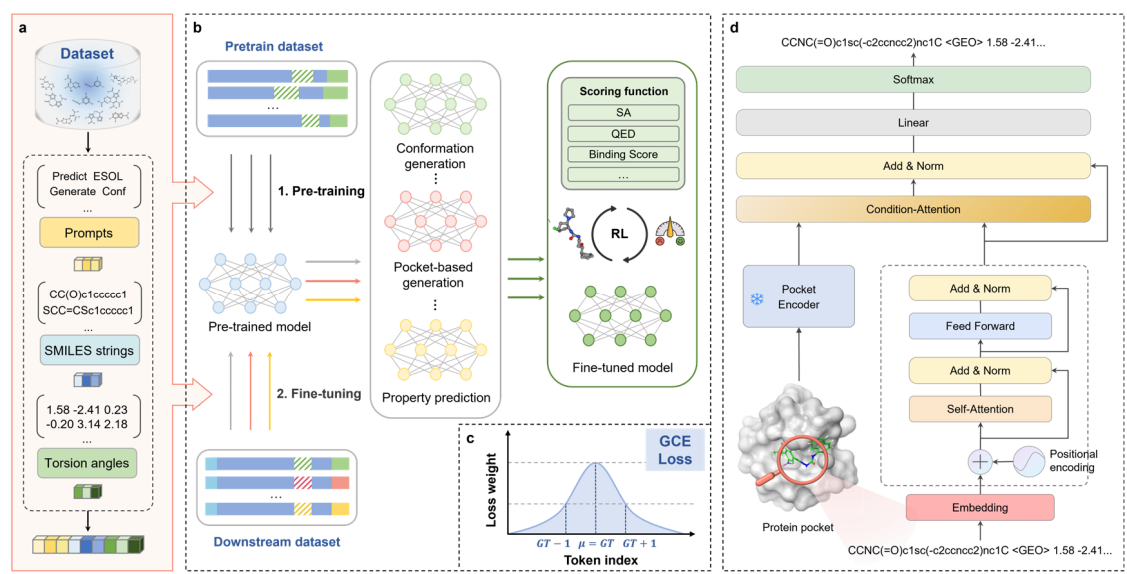

**Fig. 1 | The overview of Token-Mol.** (**a**) Data processing workflow. (**b**) The workflow of Token-Mol. (**c**) The weight allocation in the GCE loss function, where GT stands for ground truth token. (**d**) Pocket encoder and fusion block of pocket-based molecular generation.

## Results

### The overview of Token-Mol

The comprehensive workflow of Token-Mol is illustrated in **Fig. 1**. The initial phase involves pre-training on the dataset (**Fig. 1a**) through random causal masking. Subsequently, the model undergoes fine-tuning on customized datasets tailored to specific downstream tasks, including conformation generation, pocket-based molecular generation, and prediction on multiple properties (**Fig. 1b**). For regression tasks, the GCE loss function (**Fig. 1c**) is utilized during the fine-tuning process. Furthermore, the performance for specific downstream tasks can be further optimized using reinforcement learning.

The preprocessing of the pretraining dataset holds crucial significance in this context. As shown in **Fig. 1a**, a depth-first search (DFS) traversal is conducted on the entire molecule in the standard SMILES format to extract the embedded torsion angles within the molecular structure. Following this, each extracted torsion angle is assimilated as a token appended to the SMILES string. Throughout the pretraining phase, random causal masking based on causal regression is implemented. After pretraining, fine-tuning is carried out across downstream tasks. Importantly, the task prompts are specifically designed for the construction of a dialogue system, as indicated by the highlighted yellow box in **Fig. 1a**. This feature highlights a key advantage of token-only models over other large-scale models: their capability to facilitate real-time interaction. At the end of the **Results** section, examples will be presented to illustrate this particular advantage.

For the pocket-based molecular generation task, we have introduced pocket encoder and fusion block modules to better incorporate protein pocket information into the model. As depicted in **Fig. 1d**, we utilized a multi-head condition-attention mechanism to thoroughly incorporate information generated at each autoregressive step into subsequent iterations. This mechanism treats each token generated during autoregression as a prerequisite for further generation, thereby ensuring that the entire query, key, and value matrices originate from the original sequence.

One should note that, in practical scenarios, a lead compound must not only exhibit high affinity for the target but also meet a series of criteria, including high bioactivity and multiple favorable pharmacological properties. This puts higher requirements for pocket-based molecular design tasks, where the integration of receptor-ligand molecule pairs in the training dataset imposes inherent limitations. The model predominantly generates ligand molecules by utilizing information derived from the protein pocket. Consequently, the properties of these generated molecules are heavily influenced by the training data, restricting the explicitly control over their biophysical and chemical properties. These constraints are particularly evident when a precise modulation of molecular properties is desired. Token-Mol, built on an autoregressive language model architecture, where token generation aligns with actions in the RL framework, facilitates the seamless utilization of RL for optimization, thereby ensuring tailored outcomes.

**Molecular Conformation Generation**

Molecular conformation is a crucial determinant of the chemical, physical, and biological properties of molecules, underscoring its fundamental importance in structure-based drug design. The integrity and diversity of three-dimensional molecular conformations are essential for various applications in drug discovery, including three-dimensional quantitative structure-activity relationships, molecular docking and thermodynamic calculations. Traditional techniques for obtaining accurate molecular conformations, such as X-ray crystallography and nuclear magnetic resonance (NMR), are either prohibitively expensive or computationally demanding, rendering them impractical for large-scale dataset analysis. The emergence of deep geometric learning has introduced promising alternative methodologies for the generation of molecular conformations[33, 35-43].

In this study, we benchmarked our proposed approach against established baselines using widely recognized conformation generation benchmarks. We employed the dataset utilized by Zhang et al., which includes the dataset from Shi et al. (test set I) comprising 200 molecules, each with fewer than 100 conformations. It is noteworthy that this particular dataset is among the most extensively employed within the conformer generation task. On the other hand, the GEOM-Drug dataset presents a broader range of conformation counts per molecule, from 0 to 12,000. To address this variance, Zhang et al. introduced test set II[26], consisting of 1,000 randomly selected molecules with conformation counts distributed similarly to the entire dataset, ranging from 0 to 500.

Our evaluation metrics include both Recall and Precision. Recall measures the diversity of the generated conformations, while Precision evaluates the rationality of the generated conformations. We calculated the mean scores of coverage (COV) and matching (MAT) for both Recall and Precision. COV quantifies the extent to which the quantum computation conformation set covers the generated conformation set within a specified RMSD threshold, with higher values indicating better coverage. Conversely, MAT assesses the similarity between the generated conformations and the quantum mechanical-level training conformations, with lower values suggesting better performance.

**Table 1** presents the results for test set I. It indicates that Token-Mol surpasses other SOTA methods in both Precision metrics, resulting in substantial advantages. Notably, Token-Mol achieves significant superiority in the COV Precision (COV-P) metric, outperforming Tora3D by approximately 11%, underscoring the superior quality of molecules produced by Token-Mol relative to alternative methods. However, Token-Mol's generated conformations exhibit slightly lower Recall performance compared to GeoDiff and Tora3D, positioning it as the second-highest performer overall.

The findings for test set II, depicted in **Table 2**, reveal Token-Mol's exemplary performance across all assessment metrics. Remarkably, Token-Mol attains the highest performance in both Precision-based evaluation metrics, COV-P and MAT-P, surpassing other models by approximately 24% and 21%, respectively.

**Table 1 | Performance comparison of models on test set I.**

| Model | COV-R (%) ↑ | MAT-R (Å) ↓ | COV-P (%) ↑ | MAT-P (Å) ↓ |
|---|---|---|---|---|
| CGVAE | 0.00 | 3.0702 | - | - |
| GraphDG | 8.27 | 1.9722 | 2.08 | 2.4340 |

| Model | COV-R (%) ↑ | MAT-R (Å) ↓ | COV-P (%) ↑ | MAT-P (Å) ↓ |
|---|---|---|---|---|
| CGCF | 53.96 | 1.248 | 21.68 | 1.8571 |
| ConfVAE | 55.20 | 1.2380 | 22.96 | 1.8287 |
| GeoMol | 67.16 | 1.0875 | - | - |
| ConfGF | 62.15 | 1.1629 | 23.42 | 1.7219 |
| GeoDiff | **82.96**★ | 0.9525 | 48.27 | 1.3205 |
| Tora3D | 80.37 | **0.9272**★ | 62.22☆ | 1.1524☆ |
| Token-Mol | 80.65☆ | 0.9488☆ | **69.20**★ | **1.0865**★ |

★ represents the best, ☆ represents the second best.

**Table 2 | Performance comparison of models on test set II.**

| nRotb | Model | COV-R (%) ↑ | MAT-R (Å) ↓ | COV-P (%) ↑ | MAT-P (Å) ↓ |
|---|---|---|---|---|---|
| All nRotb | CGVAE | 40.06 | 1.3771 | - | - |
|  | GeoMol | 72.50 | 1.1000 | 61.15 | 1.2009 |
|  | Tora3D | 81.92 | 0.9297 | 62.16 | 1.1600 |
|  | Token-Mol | **82.34** | **0.8936** | **76.87** | **0.9107** |
| nRotb ≤ 10 | CGVAE | 42.43 | 1.3296 | - | - |
|  | GeoMol | 76.36 | 0.9380 | 57.29 | 1.1611 |
|  | Tora3D | 83.03 | 0.8704 | 63.81 | 1.0906 |
|  | Token-Mol | **83.25** | **0.8404** | **78.96** | **0.8108** |
| nRotb > 10 | Tora3D | 57.23 | 1.2455 | 29.02 | 1.5583 |
|  | Token-Mol | **65.09** | **1.1257** | **47.52** | **1.3670** |

Subsequently, we investigated the relationship between the benchmark performance and the number of rotatable bonds, as illustrated in **Fig. 2**. Our analysis reveals a clear trend: the performance across all assessment metrics declines as the number of rotatable bonds increases. This decline becomes particularly pronounced when the number of rotatable bonds exceeds 10. Notably, Tora3D exhibits a significant drop in performance when generating conformations for molecules with a higher number of rotatable bonds. In contrast, Token-Mol demonstrates substantial advantages under these conditions.

Moreover, Token-Mol demonstrates impressive speed. During our evaluation on test set I, utilizing the Tesla V100 for the generation process, Token-Mol required an average of 6.37 seconds to generate all conformations for a single molecule, compared to 8.78 seconds per molecule for Tora3D.

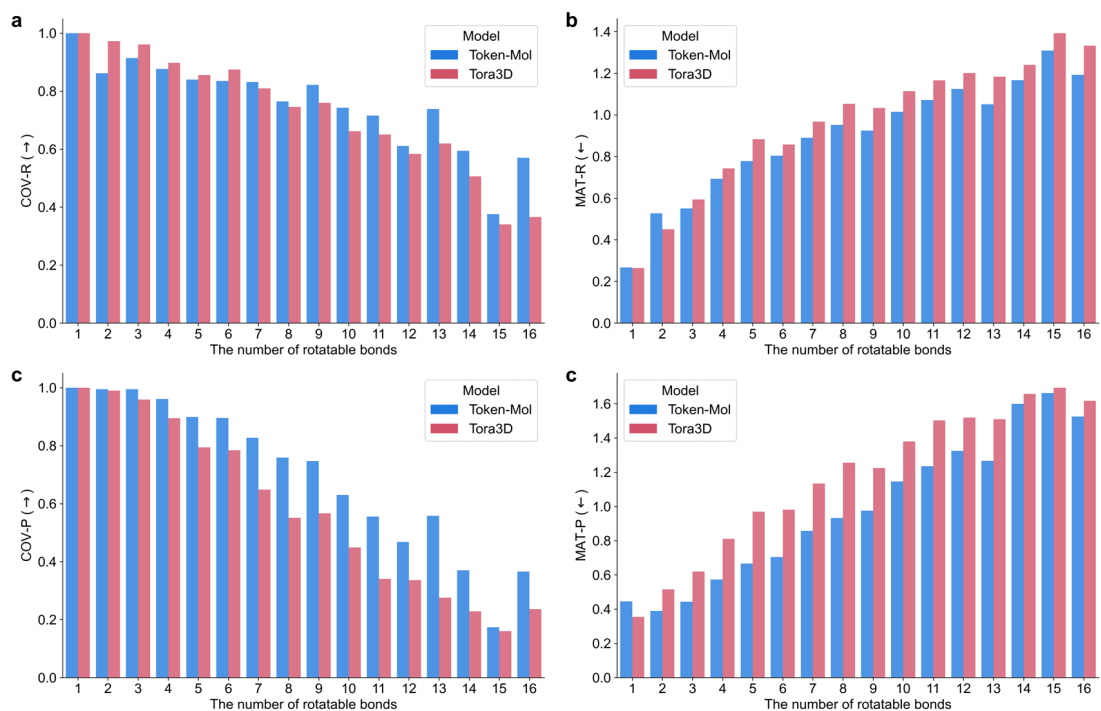

**Fig. 2 | Performance for different number of rotatable bonds on test set II.** The x-axis represents the number of rotatable bonds, and the y-axis indicates the prediction performance. **(a)** COV-R, **(b)** MAT-R, **(c)** COV-P and **(d)** MAT-P.

**Molecular property prediction**

Molecular representation is fundamental to molecular design, as it critically influences the execution of downstream tasks. In this study, we initially assessed the molecular representation capabilities of Token-Mol in the context of molecular property prediction. For a detailed description of the tasks, please refer to the Supplementary Information.

**Classification task**. For the classification task, we selected six commonly used classification datasets and compared Token-Mol against five representative baselines:

XGBoost[44] (conventional machine learning), K-Bert[45] (sequence-based model), Chemprop[46] (graph neural networks), GEM[26] (geometry-enhanced graph neural networks), and MapLight+GNN[47] (an integrated model combining traditional machine learning with graph neural networks). As outlined in **Table 3**, Token-Mol demonstrates noteworthy performance across all datasets, outperforming XGBoost and Chemprop in terms of accuracy, albeit marginally trailing behind MapLight+GNN and GEM. Notably, Token-Mol achieves state-of-the-art proficiency on single-task-focused datasets such as BBBP and BACE.

**Table 3 | Performance on different dataset for classification tasks.**

| | Classification (ROC-AUC %↑) | | | | | | |
|---|---|---|---|---|---|---|---|
| **DataSets** | **BBBP** | **BACE** | **ClinTox** | **Tox21** | **ToxCast** | **SIDER** | **Average** |
| **#Moleculars** | **2039** | **1513** | **1478** | **7831** | **8575** | **1427** | **-** |
| **#Tasks** | **1** | **1** | **2** | **12** | **617** | **27** | **-** |
| XGBoost | 0.888 ± 0.028 | 0.872 ± 0.016 | 0.863 ± 0.034 | 0.801 ± 0.061 | 0.668 ± 0.164 | 0.652 ± 0.086 | 0.791 |
| Chemprop | 0.927 ± 0.021 | 0.865 ± 0.037 | 0.877 ± 0.037 | 0.845 ± 0.015 | 0.736 ± 0.005 | 0.639 ± 0.028 | 0.815 |
| MapLight+GNN | 0.912 ± 0.026 | 0.883 ± 0.007 | 0.895 ± 0.041 | **0.865 ± 0.067** | 0.771 ± 0.156 | **0.695 ± 0.051** | 0.836 |
| GEM | 0.940 ± 0.022 | **0.898 ± 0.019** | 0.940 ± 0.026 | 0.862 ± 0.014 | **0.766 ± 0.009** | 0.670 ± 0.012 | **0.846** |
| K-Bert | **0.945 ± 0.008** | 0.879 ± 0.028 | 0.913 ± 0.046 | 0.665 ± 0.004 | 0.510 ± 0.003 | 0.608 ± 0.012 | 0.757 |
| Token-Mol | 0.934 ± 0.001 | 0.896 ± 0.015 | 0.927 ± 0.021 | 0.829 ± 0.005 | 0.746 ± 0.012 | 0.644 ± 0.020 | 0.829 |

**Regression task.** We employed a set of six regression datasets for a thorough comparison and analysis. To extend beyond established benchmarks, we introduced the token-only Regression Transformer (RT) [32], a model conceptually akin to Token-Mol, to enrich our evaluation framework. Both RT and Token-Mol fully tokenize the input and output, enabling seamless integration with foundational large models, a feature not shared by other models.

A key advantage of token-only models over traditional regression models is their ability to interface seamlessly with large models such as ChatGPT, enabling real-time interaction. However, previous models like RT have shown suboptimal performance in prediction tasks, limiting their utility for high-quality interactions. In contrast, Token-Mol treats each numerical value as a single token, rather than decomposing them into multiple tokens like RT. This approach enables one-shot prediction, thereby accelerating the prediction process. Combined with the GCE, Token-Mol achieves high-quality prediction results. This methodology allows Token-Mol to perform faster and deliver higher prediction quality.

As illustrated in **Table 4**, Token-Mol's superiority in regression tasks is evident, outperforming established benchmarks such as XGBoost, K-Bert, and token-only RT. Notably, Token-Mol consistently surpasses RT across all tasks, showcasing an average performance enhancement of approximately 30%. Particularly remarkable is Token-Mol's substantial performance boost on the Aqsol dataset, achieving an improvement of around 50%. Additionally, as depicted in **Table 5**, Token-Mol's performance closely mirrors that of graph neural network-based models on datasets with large amounts of data, such as Aqsol, LD50, and Lipophilicity. These results collectively underscore the significant potential of Token-Mol in property prediction tasks.

**Table 4 | Performance on different dataset for regression tasks.**

| | Regression (RMSE↓) | | | | | | |
|---|---|---|---|---|---|---|---|
| **DataSets** | **ESOL** | **FreeSolv** | **Lipo** | **Caco2** | **LD50** | **Aqsol** | **Average** |
| **#Moleculars** | 1128 | 642 | 4200 | 906 | 7385 | 9982 | - |
| XGBoost | 1.112 ± 0.086 | 1.958 ± 0.245 | 0.909 ± 0.032 | 0.455 ± 0.031 | 0.651 ± 0.024 | 1.540 ± 0.017 | 1.104 |

| Model | | | | | | | |
|---|---|---|---|---|---|---|---|
| Chemprop | 0.549 ± 0.028 | 1.106 ± 0.125 | 0.603 ± 0.020 | 0.429 ± 0.019 | 0.600 ± 0.021 | 1.000 ± 0.038 | 0.715 |
| MapLight+GNN | **0.529 ± 0.062** | **0.959 ± 0.278** | 0.623 ± 0.018 | 0.352 ± 0.016 | 0.600 ± 0.032 | 1.023 ± 0.031 | 0.681 |
| GEM | 0.543 ± 0.041 | 0.976 ± 0.140 | **0.584 ± 0.030** | **0.345 ± 0.038** | **0.576 ± 0.015** | **0.964 ± 0.033** | **0.665** |
| K-Bert | 0.671 ± 0.086 | 1.026 ± 0.077 | 0.641 ± 0.011 | 0.377 ± 0.022 | 0.596 ± 0.043 | 1.102 ± 0.025 | 0.736 |
| RT | 0.657 ± 0.031 | 1.389 ± 0.235 | 1.046 ± 0.528 | 0.483 ± 0.049 | 0.698 ± 0.055 | 2.344 ± 0.630 | 1.103 |
| Token-Mol (w/o GCE)* | 0.722 ± 0.022 | 1.468 ± 0.220 | 0.670 ± 0.028 | 0.441 ± 0.048 | 0.644 ± 0.025 | 1.237 ± 0.050 | 0.864 |
| Token-Mol | 0.593 ± 0.036 | 1.225 ± 0.211 | 0.645 ± 0.026 | 0.399 ± 0.010 | 0.611 ± 0.038 | 1.157 ± 0.064 | 0.772 |

*Token-Mol (w/o GCE) is the model without GCE.

**The efficiency of GCE.** Token-only generative models conventionally employ cross-entropy loss for regression tasks, but they often exhibit insensitivity to numerical values and fail to capture the relationships between them. To address this issue, we proposed the GCE loss function for regression-related downstream tasks in molecular property prediction. To assess the efficacy of GCE, we conducted ablation experiments to compare models with and without GCE (**Table 4**). Our results indicate that the absence of GCE significantly impairs Token-Mol's performance across all datasets, with an average RMSE increase of approximately 12%, underscoring the critical role of GCE in regression tasks. Compared to RT, which decomposes individual numerical values into multiple token representations, Token-Mol's one-shot approach, enhanced with GCE, demonstrates substantial improvements in both prediction accuracy and efficiency.

Despite the significant improvements demonstrated by Token-Mol compared to RT,

it still exhibits certain limitations relative to other large models based on GNN. This discrepancy is primarily due to the model's insufficient sensitivity to numerical values. Although we proposed the GCE loss function to address this issue, Token-Mol still lags behind graph neural network-based regression models. Future work will focus on enhancing the model's performance in regression tasks through approaches such as multi-task prediction and data augmentation.

**Pocket-based molecular generation**

In modern drug discovery, structure-based drug design holds paramount importance, driving researchers to rapidly identify high-affinity ligands within given protein binding pockets. Hence, pocket-based molecular generation, a method for generating potential ligands for specific pockets, not only avoids computationally intensive physical methods like traditional molecular docking but also broadens the exploration of chemical space. Consequently, it serves as a crucial downstream task to demonstrate the effectiveness of our proposed model. Our goal is to generate ligand molecules tailored to specific protein pockets. To achieve this, as illustrated in **Fig. 1d**, we amalgamated a pocket encoder and a fusion block. We utilized a pretrained encoder to characterize protein pockets, ensuring its parameters frozen during fine-tuning. Furthermore, we employed condition-attention to integrate both protein and molecule information, mirroring a prompt-like mechanism that incorporates protein pocket information into the ligand molecule generation process. The additional methodological details are outlined in the **Methods** section.

We compared our model with three popular baseline models, namely GraphBP[48], Pocket2Mol[49], and TargetDiff[50]. The first two models employ an autoregressive generative graph neural network (GNN) architecture, with Pocket2Mol introducing a geometric deep learning framework that enhances the perception of three-dimensional pocket features. In contrast, TargetDiff adopts a non-autoregressive, probabilistic diffusion model based on an SE(3)-equivariance network.

**Performance on benchmark.** We initially evaluated the generalization capability on pocket-based generation (without RL) by the following three criteria: fundamental attributes of the generated molecular sets, binding affinity towards a given pocket, and the physiochemical properties that indicate drug-likeness.

**Table 5 | Properties of the reference and generated molecules by our model and other baseline models.**

| Metric | Token-Mol | GraphBP | Pocket2Mol | TargetDiff |
|---|---|---|---|---|
| Valid | 0.973 | 0.830 | **1.000** | 0.972 |
| IntDiv | 0.849 | **0.879** | 0.812 | 0.860 |
| Simi **Ori.** | **0.132** | 0.051 | 0.097 | 0.107 |
| Simi **Training set** | **0.120** | 0.047 | 0.107 | 0.093 |
| Higher Score | **0.472** | 0.360 | 0.455 | 0.411 |

Valid: Validity of generated 3D structure, calculated as the proportion of 3D structures that can be translated into canonical SMILES; IntDiv: Internal diversity[51], an assessment of the distinctiveness of molecules within a molecular set, calculated using Tanimoto distance based on ECFP4 fingerprints[52, 53]; Simi: Similarity between two molecular sets, calculated same as IntDiv. Higher score: the average ratio of Vina score of generated molecules exceeding the original molecule within each pocket. The bolded values represent the best performers in that metric.

As shown in **Table 5**, the molecules generated by Token-Mol exhibit superior performance across the entire molecular set. In terms of validity, graph-based models tend to generate some molecules with structural flaws, leading to a decreased validity[50]. Our experiments partly confirm this issue in several graph-based models, excluding Pocket2mol. Although Token-Mol is a language model, inaccuracies in predicting token counts or values pertaining to torsion angles during autoregressive generation can yield invalid structures. Regarding internal diversity, Token-Mol demonstrates comparable performance to other models, highlighting its ability to explore chemical space under specific pocket constraints.

As to pocket-based molecular generation, binding affinity is a crucial metric. Consistent with established practices, we employ the Vina score as the proxy measure of binding affinity. In terms of affinity, the performance of Token-Mol is achieving the same level to baselines, which incorporate geometric deep learning principles. It also significantly outperforms the traditional GNN-based autoregressive generation model, GraphBP. Notably, 47.2% of Token-Mol's generated molecules exhibit superior binding affinity compared to the original molecules, outperforming all other models. Specifically, the median affinity (**Fig. 3, Table S1**) of the molecules generated by Token-Mol in the test set pockets exceeds those of the original molecules and the molecules generated by two expert models based on graph autoregression, GraphBP and Pocket2Mol, approaching the performance of the TargetDiff.

In comparison with other models, Token-Mol (without RL) exhibits a Vina score distribution that closely aligns with the true distribution observed in the test set, indicating that its capability in accurately capturing the actual data distribution. Conversely, Pocket2Mol and TargetDiff generate molecules with notably broader Vina score distributions that deviate significantly from the test set distribution. Although these models can generate molecules with lower Vina scores, many of them are hallucinated[54], displaying low Vina scores but containing obvious structural anomalies that render them unsuitable as drug candidates in reality. These concerns are further explored in the section **Pocket-based generation on real-world targets**.

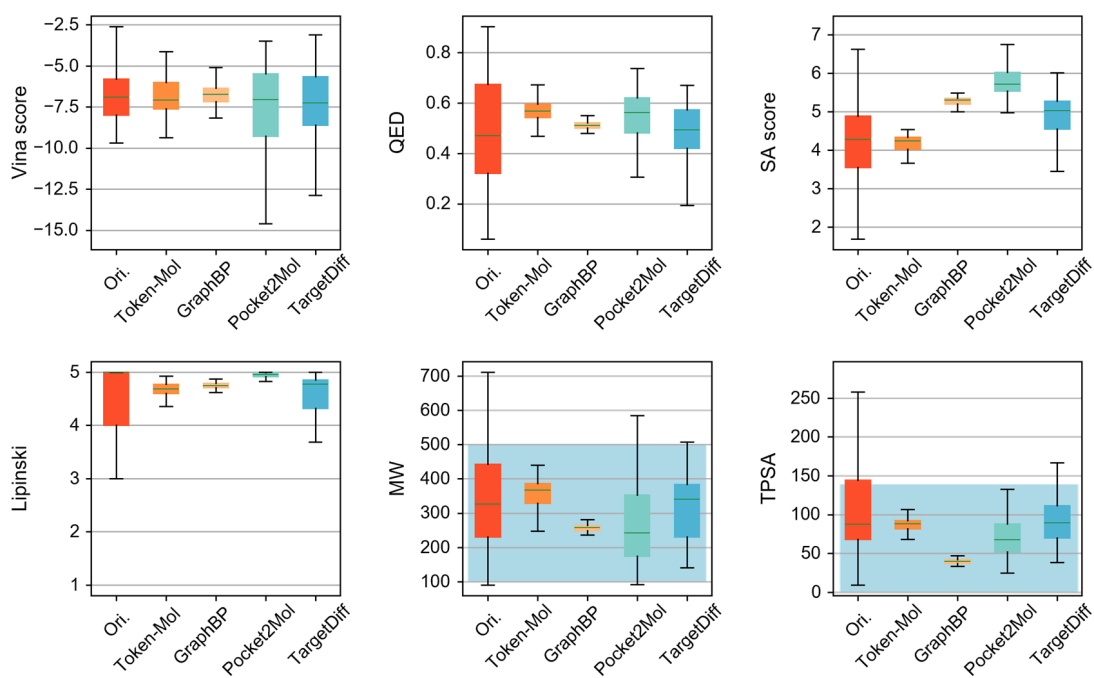

**Fig. 3 | Comparison of the average molecular properties distribution in the test set pockets for the molecules from original test set or generated by Token-Mol and**

**other baseline models.** Vina score: the binding energy of ligands to protein pockets by using QVina2[55]. QED: quantitative estimation of drug-likeness[56]. SA score: synthetic accessibility score[57]. Lipinski, the number that satisfies Lipinski's rule-of-five[58]. MW: molecular weights, the optimal range is between 100 and 600[59]. TPSA: topological polar surface area[60], the optimal range is between 0 and 140[61]. The green bars stand for the median values, and the light blue shading represents the optimal range for that metric.

Physiochemical properties of molecules play a pivotal role in drug-likeness of drug candidates. In this regard, Token-Mol outperforms other models in generating molecules with better QED and SA Score, exceeding 5~10% benchmarks, thereby demonstrating its proficiency in creating more drug-like molecules. As for Lipinski's rule metrics, taking original molecules as reference, its overall performance remains exceptional. Concerning molecular weight, Token-Mol's distribution aligns closely with the original molecules, staying within the optimal range, whereas GraphBP and Pocket2Mol have relatively lower median molecular weights, accounting for their ease in satisfying Lipinski's rule. The metric TPSA determines the oral bioavailability and membrane permeability of molecules[61], with values below 140 Å$^2$ for cell membrane traversal and below 90 Å$^2$ for blood-brain barrier penetration. The TPSA distribution of molecules generated by Token-Mol falls within the range of 70-100 Å$^2$, which is more reasonable compared to other baseline models, suggesting superior absorption and potential for further drug discovery in central nervous system diseases[62].

Beyond metrics such as binding affinity and molecular properties, the fidelity of torsion angles within generated molecules needs to be considered. Torsion angles will be used as an indicator to evaluate the reasonableness of the initial conformation. Molecule with torsion angle distribution closer to that of the ground truth molecule suggests that its conformation is more likely to be closer to the real molecule and does not violate inherent physical constraints. Moreover, excessively twisted torsion angles in the initial conformation can induce the conformation's energy to become trapped in local minima during molecular docking, making it difficult to escape and causing deviations in the docking results. Therefore, reasonable torsion angles are also beneficial for virtual screening based on docking.

Our analysis involved the examination of torsion angles within the test set (**Fig. S1**). Subsequently, we curated a subset of torsion angles, characterized by their abundance and non-random distribution, enabling an in-depth comparative analysis. Jensen-Shannon divergence is used to assess the disparity between the torsion angle distributions in the test set and those of the molecules generated by the models (**Table 6**).

**Table 6 | Jenson-Shannon divergence between the test set and molecules generated by Token-Mol and baseline models.**

| | Jenson-Shannon Divergence ($\times 10^{-1}$) ↓ | | | |
|---|---|---|---|---|
| | Token-Mol | GraphBP | Pocket2Mol | TargetDiff |
| CNCC | **0.240** | N/A | 1.749 | 1.637 |
| C^CCC | **0.281** | N/A | 1.385 | 1.496 |
| CCCC | **0.326** | N/A | 1.672 | 1.675 |
| NCCC | **0.421** | N/A | 2.269 | 1.767 |
| NCC^C | **0.427** | N/A | 2.173 | 1.644 |
| COC^C | **0.432** | N/A | 1.372 | 1.545 |
| CCC=O | **0.336** | N/A | 1.496 | 1.547 |
| CNC=O | **0.405** | N/A | 1.517 | 1.581 |
| CNC^C | **0.393** | N/A | 1.676 | 1.406 |
| C^CC^C | **0.552** | N/A | 1.397 | N/A |
| Avg. | **0.381** | N/A | 1.671 | 1.589 |

N/A means this type of torsion angle did not appear in the molecules or is not detected due to the unreasonable conformation of the molecule. Symbol in the bond types: "=" represents double bond, "^" represents aromatic bond.

Furthermore, we calculated the average molecular generation time for each model. A faster generation speed indicates the ability to explore the chemical space under pocket constraints as much as possible in a limited timeframe, thereby accelerating the drug discovery process. This efficiency also reduces the demand for computational resources, enabling researchers with limited resources to effectively utilize the model.

To ensure a fair comparison, we measured the cumulative time spent by each model in sampling pockets and generating molecules until an output file (in sdf/mol2 format) was obtained. As shown in **Table 7**, Token-mol emerged as the fastest among the various models. When compared to models utilizing geometric deep learning frameworks, Token-Mol demonstrated a remarkably higher generation speed, averaging approximately 35 times faster for individual molecules. This efficiency stems from the different methodologies adopted by competing models. For instance, Pocket2Mol necessitates extensive sampling of molecular objects, excluding invalid or duplicated molecules to maintain diversity and validity. Similarly, TargetDiff requires performing thousands of rounds of sampling on the atoms within the pocket before molecular generation, ensuring high-quality outputs but significantly impeding the generation process for both models.

Furthermore, we evaluated the generation capabilities of the Token-Mol model on a cost-effective GeForce RTX4070 GPU with 12GB VRAM. The results indicate that the model can achieve a maximum batch size of 70, generating molecules within a given pocket at an average rate of 0.424s per molecule. This highlights the feasibility of deploying the model on consumer-grade computers for rapid inference.

**Table 7 | Generation speed between different models on Tesla V100 GPU.**

|  | Token-Mol | GraphBP | Pocket2Mol | TargetDiff |
|---|---|---|---|---|
| Times Spent (s/molecule) | **0.365** | 0.750 | 13.103 | 12.971 |

**Drug design for real-world targets**

To evaluate the generatability of models in designing drug candidates for real-world therapeutic targets, we selected 8 targets from three important protein families, namely kinases, G-protein coupled receptors (GPCRs) and viral proteins, which had been widely studied in structure-based drug design[63-65] and molecular generation[66-69]. Specially, our selection includes a unique dimeric pocket from Programmed cell death 1 ligand 1 (PD-L1), aiming to explore the models' capability in designing small-molecule modulators for protein-protein interactions.

To mimic a realistic drug discovery scenario, we generated an equal number of molecules across diverse targets, performed molecular docking, and calculated the molecular properties such as QED and SA Score for each molecule. Our goal was to produce potent "drug-like" molecules that possess high affinity to the target, excellent drug-likeness, and favorable synthetic accessibility. To this end, we set criteria that a "potent drug-like" molecule should simultaneously satisfy a Vina Score lower than the average Vina Score of reference molecules from its corresponding target (**Table S5**) a QED of at least 0.5, and an SA Score not exceeding 5.0.

As shown in **Table 8**, our approach achieved the highest ratio in six out of eight targets and the second-best result on the DDR1 and PD-L1 pocket. Specifically, For the targets including kinases except for DDR1, GPCRs and viral protein 3CLpro, the Token-Mol model significantly outperformed other models, with over 15% of the generated molecules meeting our predefined criteria for "potent drug-like" molecules in average, which indicating its potential to generate reasonable molecules for real-world therapeutic pockets. For the PD-L1 dimeric protein pocket, Token-Mol displayed a comparable performance than other models. Further analysis of the distributions of Vina Scores and QED (**Fig. S2**, **Table S2**) for the molecules generated by those models reveal that Token-Mol not only produces molecules with high affinity but also ensures they possess desirable properties. This aligns with the results for our test sets, suggesting that our model is capable of identifying promising lead compounds in real-world drug discovery scenarios.

**Table 8 | The ratio of the "potent drug-like" molecules generated by each model**

**for each target.**

| | "Potent drug-like" molecules (%) | | | |
|---|---|---|---|---|
| Target | Token-Mol | GraphBP | Pocket2Mol | TargetDiff |
| CDK2 | **9.091** | 0 | 0 | 2.174 |
| AKT1 | **24.138** | 1.176 | 0 | 2.174 |
| EGFR | **12.037** | 0 | 0 | 0 |
| DDR1 | 3.704 | 0 | 0 | **5.128** |
| ARA2A | **15.455** | 1.176 | 0 | 14.737 |
| ADRB2 | **13.889** | 1.136 | 0 | 1.099 |
| 3CLPro | **25.225** | 1.205 | 0 | 6.410 |
| PD-L1 | 59.259 | 26.087 | **85.841** | 9.589 |
| Mean | **20.350** | 3.848 | 10.730 | 5.164 |

Furthermore, to evaluate the gains of RL within this model framework, we selected cyclin-dependent kinase 2 (CDK2), representing kinases, and the adenosine A2A receptor (ARA2A), representing GPCRs, as two moderately performing targets from two significant families. As shown in **Fig. 4**, the molecules generated by Token-Mol exhibit favorable drug-likeness, synthesizability, and promising affinity within the target pockets of two proteins that exhibit significant structural and functional differences. These molecules possess more rational structures compared to those generated by other models and display distinct scaffolds between the two different targets.

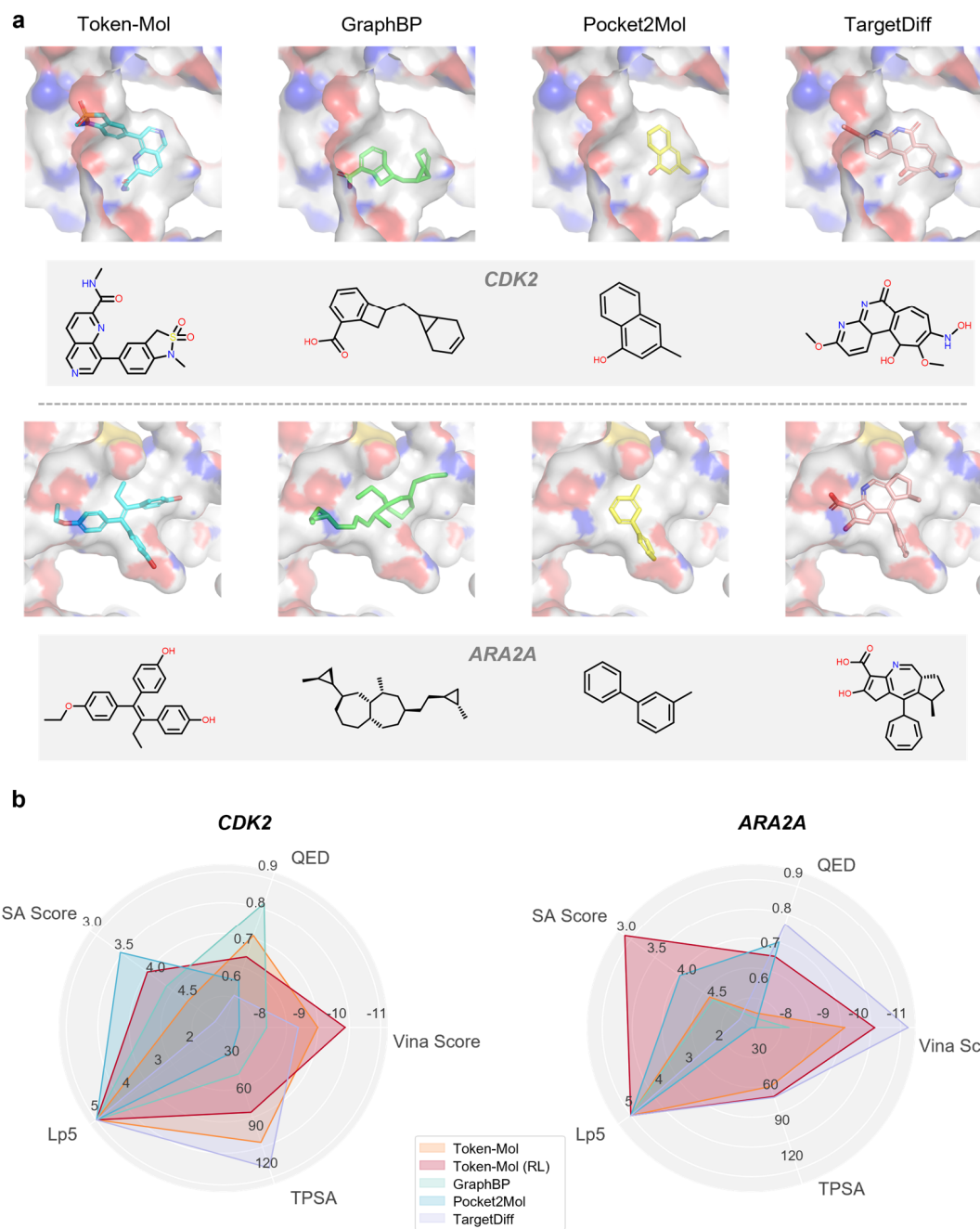

**Fig. 4 | Evaluation on real-world targets.** Comparison between (**a**) structures and binding modes, and (**b**) related molecular properties of "Drug-like" molecules with the highest affinity generated for CDK2 and ARA2A by the Token-Mol and baseline models. The detailed information of molecular properties is presented in **Table S3**.

Among the molecules generated by other baseline models, those produced by GraphBP exhibit distorted conformations, while those by Pocket2Mol are simple aromatic ring derivatives and exhibit minimal differences between the two targets.

Similar phenomena can also be observed in the molecules generated by the aforementioned two models in other use cases (**Fig. S3**). As for the molecules generated by TargetDiff, while they demonstrate favorable results in terms of QED and Vina score, it is noteworthy that molecules for two targets contain tricyclic scaffolds and 7-membered cyclic groups, which are challenging to synthesis. This can explain why the molecules generated by TargetDiff have higher Vina scores in the former test, as these groups with large volume occupy as much space as possible within the pocket, creating more hydrophobic contacts. The predicted binding modes further confirm this, as the molecules generated by Token-Mol fit the shape of the pocket cavity more precisely, whereas those generated by other baseline models only occupy part of the pocket cavity.

To further demonstrate Token-Mol's capability to generate molecules that resemble real-world ligands, we chose ARA2A as the target to analyze the similarity between ligands and generated molecules. We selected several molecules from Token-Mol and TargetDiff that exceeded the average Vina score and QED thresholds of reference molecules (**Table S5**) against ARA2A for display. In **Fig. 5**, we present six real ligands of ARA2A, including agonists and antagonists. Notably, adenosine, the leftmost ligand, serves as the natural ligand of ARA2A and contains a purine scaffold. The other discovered ARA2A ligands all possess a nitrogen heterocyclic core as their scaffold, similar to purine, which can be monocyclic, bicyclic, or tricyclic[70]. From the perspective of medicinal chemists, for the antagonists, which are majority of ARA2A ligands, their structure-activity relationship (SAR) and co-crystallized structures indicate that, in addition to the nitrogen heterocyclic scaffold, there are aromatic rings such as furan, thiophene, or benzene directly connected to the scaffold or located one or two carbon atoms away from it[71]. These aromatic groups can fit into the internal space of the target pocket and engage in π-π stacking interactions[72], thereby enhancing the ligand's affinity for the receptor and enabling strong competition with the natural ligand.

Among the molecules generated by Token-Mol, it can be observed that most contain monocyclic or bicyclic nitrogen heterocyclic scaffolds resembling real-world ligands, whereas those generated by TargetDiff differ significantly from real-world ligands. Furthermore, these molecules with nitrogen heterocyclic scaffolds possess aromatic rings, such as benzene or pyrazole, directly connected to the scaffold, consistent with the aforementioned SAR. This suggests that these molecules may conform to the SAR of ARA2A antagonists and exhibit good affinity for the target.

Simultaneously, we calculated the similarity of Bemis-Murcko scaffold[73] and Fréchet ChemNet Distance[74] (FCD) between the molecules generated by Token-Mol and other baseline models and the reference molecules (**Table S4**). The results demonstrate that Token-Mol generates molecules with higher similarity to the real-world ligands across all tested targets compared to other baseline models. Importantly, this similarity remains within an acceptable range to ensure molecular novelty, further supporting our conclusion that Token-Mol can generate molecules that closely resemble real-world ligands.

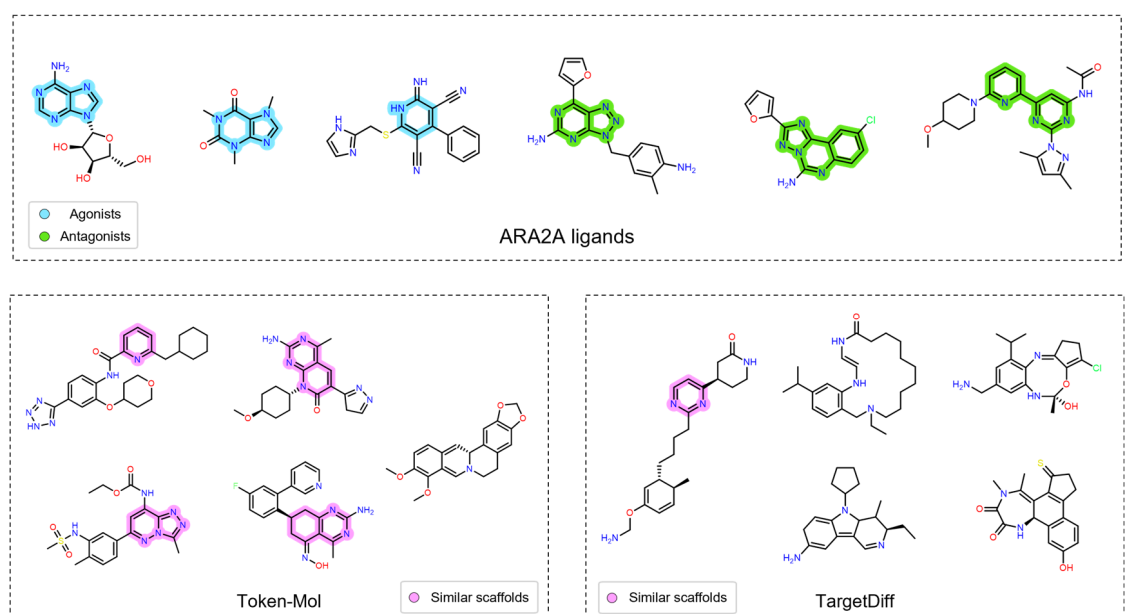

**Fig. 5 | Comparison between the real-world ligands of ARA2A and molecules with generated by Token-Mol and TargetDiff.** The scaffolds of the real-world ARA2A ligands and those of molecules similar to the real-world ligands generated by the models are highlighted.

**Further optimization with reinforcement learning**

The content above has already demonstrated that Token-Mol can generate molecules with high drug-likeness, ease of synthesis, and rational structures in the pockets of given targets. However, when compared to the expert models like TargetDiff for pocket-aware generation, the molecules generated by Token-Mol still exhibit lower affinity to pockets from the test set and selected real-world targets (**Tables S1 and S2**). To address this, we have introduced a reinforcement learning approach to optimize the affinity of generated molecules for specific target pockets. Within this framework, constraints such as conformational clash and drug-likeness are enforced to ensure that the molecules maintain desirable properties. This strategy aims to maximize affinity for target pockets while preserving the excellent molecular properties demonstrated by the model, as detailed in the **Methods** section. Notably, optimizing geometric graph-based models such as TargetDiff is challenging due to their high complexity, and RL has not yet been applied to these models for 3D pocket-based molecular generation.

We conducted a total of 1,000 steps of reinforcement learning optimization on CDK2 and ARA2A, the two targets used for demonstration in the previous section. Throughout the reinforcement learning process, we recorded the average values of key metrics at each step (**Fig. 6a**). It can be observed that during the training of the two targets, the reward score essentially converged within 1,000 steps, indicating the stabilization of the agent model's training. Regarding our primary optimization objective, the Vina score, the average value is optimized from around -8 to approximately -9.5, with no significant oscillations observed after convergence. As for QED, which serves as a constraint condition, although the reward term in the reward function is binary rather than positively correlated with QED, it was found that the QED value initially increased and then converged as the reinforcement learning steps increased for both targets, suggesting that QED is also optimized under the set reward function. Although different trends were observed in the SA score during the reinforcement learning process for the two targets, the results remained below the threshold of 5, consistent with our previous tests (**Fig. S4**) that focused solely on affinity optimization. These trends demonstrate that our model can achieve optimization in molecule generation tasks for specific target pockets through reinforcement learning under constraints.

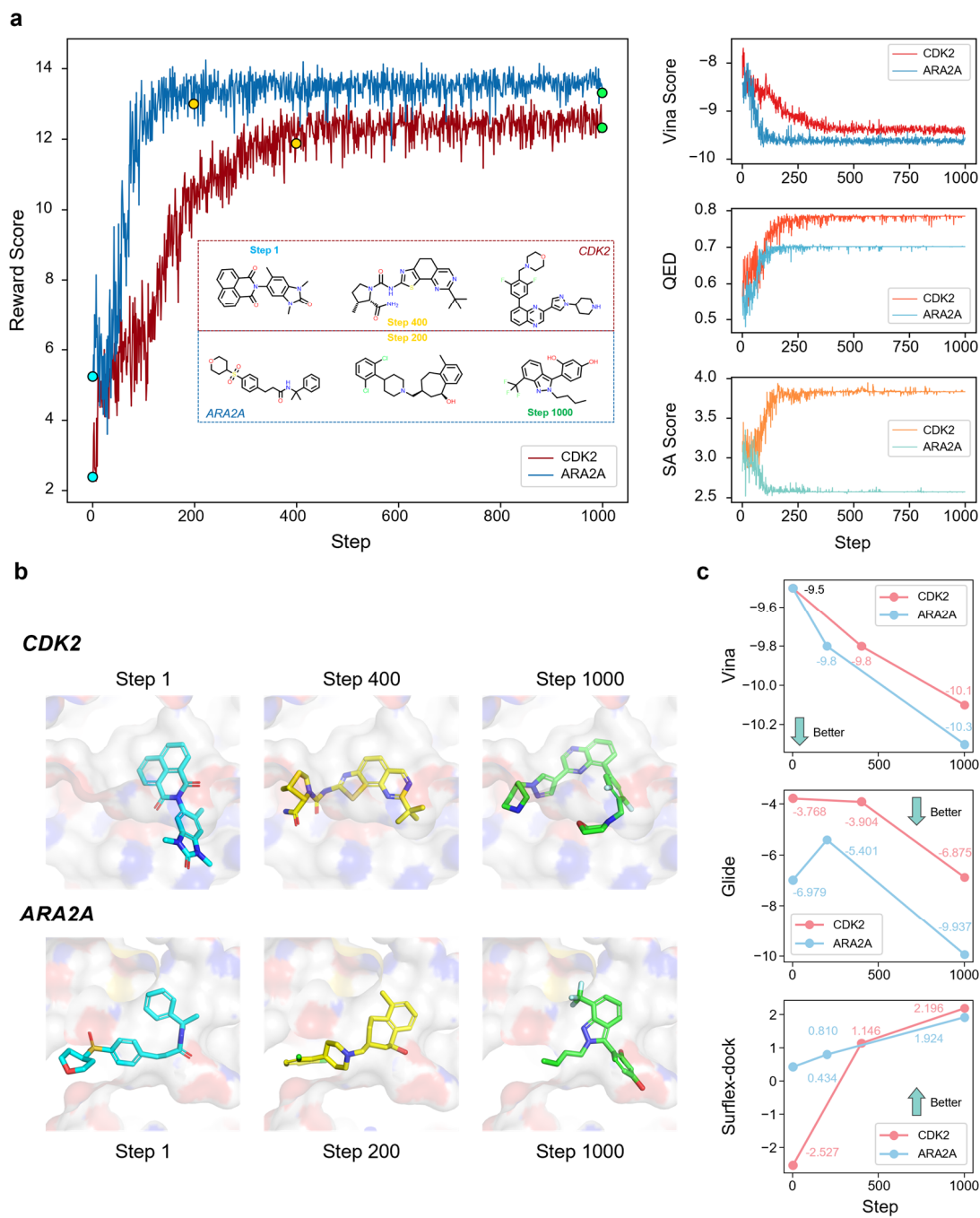

**Fig. 6 | Molecular performance in RL process.** (**a**) Key metrics such as reward score, Vina score, QED and SA score during the process of RL. The 2D structure of molecules from the different stages in the RL are also displayed. (**b**) The binding modes of selected molecules predicted with QVina2. (**c**) Change trends in the predicted affinities of selected molecules for their respective targets using different docking methods.

Additionally, we selected molecules from the first step, the last step, and step before the convergence of the reward score, showcasing the molecule with the highest affinity in those steps (**Fig. 6a**). From the perspective of specific molecules, it is obvious that the molecular scaffolds undergo substantial changes at different training stages, and the occurrence of unreasonable structures such as tricyclic structures or seven-membered ring groups also decreases. For their binding modes (**Fig. 6b**), it can be seen that the scaffold gradually fits into the pocket, which explains the gradual improvement of the Vina score during the training process.

Furthermore, to reduce the bias introduced by a single docking method, we conducted additional docking tests using Glide[75] and Surflex-dock[76] for the molecules presented in **Fig. 6c**, and the results demonstrate that reinforcement learning indeed optimized the molecules' affinity for the targets. In the case of CDK2, the docking scores (i.e., the predicted affinities) obtained from all three methods improved as the training steps increased. In ARA2A, a similar trend was observed, with the exception of Glide. Overall, in both targets, the molecules obtained after training convergence achieved the best results across all three docking methods, further confirming the capability of reinforcement learning to optimize the affinity of generated molecules for the target pockets.

**Chat to Token-Mol**

Token-Mol exhibits significant potential compared to existing pre-trained large models, particularly in its ability to integrate seamlessly with general-purpose large models and facilitate unrestricted dialogue. This integration leverages techniques from various large language models, including prompt learning, mixture of experts (MoE)[77], and retrieval-augmented generation (RAG)[78].

To illustrate this capability, we present several straightforward dialogue use cases. By employing prompt learning, we can control the execution of tasks such as property prediction mentioned in this study. Initially, we insert specific prompts, such as "Predict ESOL" to fine-tune the model. As shown in **Fig. S5**, this enables direct interaction with the model post-prompting, allowing users to request predictions of different molecular properties. In this example, we queried various properties of different molecules, and

Token-Mol successfully provided the corresponding predictions. This demonstrates potential of Token-Mol for engaging in meaningful dialogues with chemists. Users may provide molecular conformations, but since Token-Mol can generate the corresponding conformations, the final output will include only the predicted target properties.

Additionally, future iterations can incorporate RAG. When querying Token-Mol about a specific property of a molecule, the system employs vector search based on embeddings to convert the query into a vector. This vector is then matched with highly relevant vector descriptions from a database to provide contextual information. The query, along with the retrieved context such as spatial structure information and other relevant properties, is then input to Token-Mol, which then generates the answer.

The aforementioned example highlights the unique of token-only models to seamlessly integrate with general models, a capability that is not exhibited by other types of models.

## Discussion

This study proposes Token-Mol, the inaugural token-only extensive pre-trained language model tailored for drug design. Rooted in the GPT framework, Token-Mol integrates random causal masking to enhance its flexible applicability across diverse drug design tasks. Additionally, we propose the Gaussian cross-entropy loss function to foster improved acquisition of continuous spatial information throughout model training, thereby notably reinforcing its performance in regression tasks. Furthermore, through the integration of RL, Token-Mol achieves expedited optimization towards predetermined objectives in specific tasks, aiming to achieve desired outcomes efficiently. To substantiate these capabilities, we conducted assessments across three pivotal drug design tasks.

In pocket-based generation task, Token-Mol achieve results close to expert models in the pocket-based generation task and obtain optimal results in terms of drug-likeness and synthesizability of molecules. Benefit from the rapid inference of the language model, Token-Mol can generate molecules within the pocket in a shorter time.

Additionally, tests on specific real-world targets have also demonstrated that our model can obtain molecules with excellent affinity, drug-likeness, and synthesizability under various conditions simulating real-world virtual screening with a higher proportion. For specific optimization goals in the specific targets, we preformed reinforcement learning, and the results also proved that Token-Mol can achieve optimization under constraint conditions, demonstrating the broad application potential of our model.

We subsequently evaluated its capability in molecular conformation generation. Token-Mol demonstrated superior performance relative to other SOTA models, exceeding their performance by approximately 24% in COV-P and 21% in MAT-P. Notably, Token-Mol exhibited improved efficacy in molecules with a higher number of rotatable bonds.

Lastly, we assessed its performance in molecular property prediction tasks. Leveraging the advantages of the Gaussian cross-entropy loss function, Token-Mol demonstrated accuracy on par with state-of-the-art models. In regression tasks, Token-Mol outperformed the token-only model RT by approximately 30% and surpassed existing sequence-based methods, approaching the performance of GNN-based methods.

Meanwhile, Token-Mol demonstrates a remarkable ability to simplify complex problems, capitalizing on the inherent advantages of large language models. This proficiency is particularly pronounced in sophisticated tasks such as pocket generation, where Token-Mol achieves an optimal balance of speed and efficacy. Notably, in comparison to the state-of-the-art model TargetDiff, Token-Mol's inference speed is 35 times faster.

While Token-Mol demonstrates significant potential, several areas require further enhancement. In this study, we evaluated its performance on only three representative downstream tasks, leaving many others unaddressed. The molecular diversity within the pre-training data is also limited. Future research will focus on optimizing Token-Mol by expanding the training dataset and developing specific components tailored to particular downstream tasks. Comprehensive evaluations across a broader range of drug design tasks will be conducted. Additionally, we aim to integrate Token-Mol with general artificial intelligence models, utilizing techniques from various large language models such as prompt learning, MoE, and RAG. This integration will facilitate direct interaction between researchers and Token-Mol through conversational interfaces, enhancing its role as a research assistant.

In summary, this study presents a token-only foundational model for drug design, introducing the initial version of Token-Mol. Its development offers a novel approach towards unifying AI drug design models, paving the way for comprehensive drug design using a single foundational model.

## Methods

### Model architecture

**Backbone.** Token-Mol is structured with 12 layers of Transformer decoders, each equipped with 8 attention heads. Employing autoregressive approach, Token-Mol predicts both the 2D and 3D structures of molecules while explicitly representing them. To ensure data integrity during autoregressive training and inference, masking matrices are employed to conceal unencoded segments, thus preventing information leakage. The multi-head attention mechanism, integral to the Transformer architecture, empowers Token-Mol to simultaneously attend to diverse subspaces of the input, facilitating the capture of richer information. Within this mechanism, each attention head learns a unique set of weights to compute attention scores for different positions in the input sequence, facilitating the calculation of the input sequence's representation. By harnessing parallel computation across multiple attention heads, Token-Mol gains the capacity to interpret the input sequence from various perspectives, consequently enhancing its representational capability and generalization performance. The attention mechanism is shown in Equation 1:

$$Attention(Q, K, V) = \text{softmax}\left(\frac{QK^T}{\sqrt{d_k}}\right)V. \qquad (1)$$

where $Q$, $K$, and $V$ represent the query, key, and value matrices, respectively, and $d_k$ is the dimension of $K$.

To indicate the beginning or end of the sequence during sampling, it is necessary to define a start token and an end token, denoted as "<|beginoftext|>" and "<|endofmask|>", respectively. During the training, the "<|beginoftext|>" token is concatenated to the sequence as the input. The objective during the training phase is to minimize the negative log-likelihood, as shown in Equation 2:

$$\mathcal{L} = -\sum_{i=1}^{n} \log p(x_i | x_{<i}). \tag{2}$$

During the generation phase, peptides strings are generated using an autoregressive approach based on smiles, which are then concatenated together as shown in Equation 3:

$$p(x) = \prod_{i=1}^{n} p(x_i | x_{<i}). \tag{3}$$

**Gaussian cross-entropy (GCE) loss function.** Language models commonly employ the cross-entropy loss function as their primary loss function. The cross-entropy loss function is generally utilized to quantify the disparity between the probability distribution produced by the model and the actual labels. Assuming a classification problem, for each sample, the model outputs a probability distribution indicating the likelihood of the sample belonging to each class. The genuine labels, on the other hand, are one-hot encoded vectors representing the class to which the sample belongs. The cross-entropy loss function is employed to measure the dissimilarity between the probability distribution produced by the model and the genuine labels. In the context of language models, the specific equation for calculating the cross-entropy loss function is as follows:

$$\mathcal{L} = -\frac{1}{m}\sum_{i=1}^{m}\sum_{j=1}^{n} y_{ij} \log q(x_{ij}). \tag{4}$$

Here, $m$ represents the batch size, and $n$ denotes the length of each data point. $y_{ij}$ signifies the $j$-th element of the true label for the $i$-th data point (taking values of 0 or 1), and $q(x_{ij})$ represents the $j$-th element of the probability distribution output by the model. A lower cross-entropy loss indicates a closer resemblance between the model's output probability distribution and the true labels, thereby reflecting better model performance.

However, the conventional employment of the cross-entropy loss function is primarily confined to discrete category prediction tasks, rendering it inadequate for continuous value prediction endeavors such as regression. In our investigation, we encounter a spectrum of tasks encompassing both classification, exemplified by SMILES strings, and regression, including torsion angles and molecular property prediction. In response to this challenge, the regression transformer disassembles each digit of continuous numerical values into distinct tokens and incorporates specialized numerical embeddings. Nonetheless, their methodology does not fundamentally rectify the issue, as it neglects to facilitate the model's comprehension of the relative magnitude relationships inherent in numerical values. Notably, the model uniformly assigns loss values in the event of inaccurate predictions, irrespective of the predicted token. For instance, if the label denotes a torsion angle of $\pi$, erroneous predictions of 3 or 0 result in identical loss values.

To surmount this constraint, we propose the GCE loss function tailored specifically for regression tasks. As shown in **Fig. 1**, for each prediction, we construct a Gaussian distribution centered around the label's value, thereby adjusting the probabilities of surrounding tokens from their original values of 0 to correspond with the Gaussian distribution. Consequently, in Equation 5, where $p(x_{ij})$ is initially denoted as either 0 or 1, we modify it to signify a Gaussian distribution centered around the label's value, thereby effectively mitigating the issue. The GCE loss function is defined as:

$$\mathcal{L} = -\frac{1}{m}\sum_{i=1}^{m}\sum_{j=1}^{n}\frac{1}{\sqrt{2\pi\sigma^2}}e^{-\frac{(x_{ij}-y_{ij})^2}{2\sigma^2}}\log q(x_{ij}). \tag{5}$$

Through the implementation of this configuration, tokens in proximity to the label are allocated greater weights, whereas tokens distanced from the label receive diminished weights. This methodology facilitates the comprehensive learning of relationships between numerical values by the model.

**Pocket encoder and fusion block.** We utilized the protein pocket encoder trained by Odin et.al[67], maintaining its parameters frozen throughout the training process. To merge the information derived from the pocket encoder with the existing molecule information within the model, we employed a multi-head condition-attention mechanism. Diverging from traditional cross-attention mechanisms, our approach involved the adoption of a multi-head condition-attention mechanism to fully integrate the information generated at each autoregressive step into subsequent generations. This mechanism treats each token produced during autoregression as a prerequisite condition for iterative generation. Consequently, the entire query, key, and value matrices stem from the original sentence itself. Particularly, as shown in **Fig.1c**, this condition-attention fundamentally regards protein information as prompt data, enabling the model to analyze the interaction between protein information and previously generated tokens.

**Reinforcement learning.** REINFORCE[79], an RL algorithm based on policy gradients, utilizes Monte Carlo methods to determine the optimal policy, and it has been applied in various molecular generation methods[80] [81-83]. In this work, we used the REINFORCE algorithm to optimize the model. We aim to optimize the pre-trained model parameter $\theta$ for the task of generating molecular sequences, so that the optimized model can generate molecules with desired properties, as shown in Equation 6:

$$\theta' = argmax_\theta \left( E_{\pi_\theta}(G(\tau)) \right). \quad (6)$$

The presented formula elucidates the policy $\pi_\theta$ as contingent upon model parameters $\theta$, with $\tau$ delineating a trajectory spanning states $s_t$ and actions $a_t$ from the initial time step $t = 0$ to the terminal step $t = T - 1$. Concomitantly, the reward at each juncture t within the trajectory is designated as $r(s_t, a_t)$. Equation 7 concisely portrays the aggregate reward accumulation from time step t to the final state, encapsulating the core essence of the trajectory's reward accumulation dynamics.

$$G(s_t, a_t) = \sum_{k=t+1}^{T-1} r(s_k, a_k). \quad (7)$$

According to the REINFORCE, the objective function can be derived as follows:

$$J(\theta) = E_{\pi_\theta}\big(G(s_t, a_t)\big) = \left(\sum_{t=0}^{T-1} \log \pi_\theta(a_t|s_t)\, G(s_t, a_t)\right). \tag{8}$$

Within the molecular generation realm, computing $G(s_t, a_t)$ for each step in a trajectory, corresponding to incomplete molecules, proves impracticable given the inability to reliably estimate the total molecule score from its constituent fragments alone. This scenario converges with the sparse reward paradigm prevalent in reinforcement learning. To surmount this challenge and enable the deployment of the REINFORCE algorithm in this context, we advocate for equating the complete molecule score with the score at each step, thereby reformulating $J(\theta)$ as:

$$J(\theta) = G(\tau) \left(\sum_{t=0}^{T-1} \log \pi_\theta(a_t|s_t)\right). \tag{9}$$

Finally, the refinement of $J(\theta)$ is accomplished via gradient descent optimization.

**Reward function**. To optimize affinity, the reward function is designed to prioritize molecules that meet a promising Vina score. Molecules that exceed the affinity threshold and comply with the QED constraints receive additional rewards. Molecules that do not meet the affinity threshold or are non-compliant are penalized. Thus, the reward function $R(m)$ is described as Equation 10:

$$\mathcal{R}(m) = \begin{cases} \omega \cdot (Vina(m) - init + 0.1) + \theta_{qed}, & \text{if } Vina(m) \leq init \\ 0.1, & \text{if } Vina(m) > init \\ 0, & \text{if } invalid \end{cases} \tag{10}$$

where $m$ is molecule; *Vina(m)* represents Vina score, where a smaller value is preferable; *init* is the threshold value of Vina score, which is set as -8. To avoid the issue of sparse rewards, we have imposed a reward weight $\omega$, set as 5, and a proper penalty term set as 0.1 for molecules which do not meet the threshold of Vina score. $\theta_{qed}$ is a reward term for molecules that comply with the restraint of QED, describe as Equation 11:

$$\theta_{qed} = \begin{cases} 1, & QED \geq 0.5 \\ 0, & QED < 0.5 \end{cases} \tag{11}$$

**Random causal masking**

The conventional left-to-right causal masking method exclusively relies on the context preceding the generated tokens, thereby proving inadequate for accomplishing the infilling task. To enhance the adaptability to a wider array of downstream tasks, we opted to train it using random causal masking[84, 85] in lieu of the left-to-right causal masking.

Throughout the training process, we commence by sampling the number of mask spans from a Poisson distribution centered around a mean of 1, while enforcing a limit that confines the count of mask spans within the range of 1 to 6. Following this, we employ random sampling to establish the length of each span. The locations of the masks are identified using placeholders denoted as "<|mask:k|>", with "k" signifying the index of the specific mask span. Subsequently, the content subjected to masking is affixed to the sequence's end, preceded by the "<|mask:k|>" prefix. In the inference phase, a sequence incorporating placeholders "<|mask:k|>" is presented as the contextual input, complemented by the addition of "<|mask:k|>" at the sequence's conclusion to steer the model's generation of content for the "<|mask:k|>" segments.

**Benchmark**

**Molecular conformer generation.** COV and MAT scores are fundamental metrics utilized as benchmarks in the Conformer generation task, extensively employed across conformer generation endeavors. COV and MAT metrics are further categorized into Recall and Precision measures. Recall is defined as:

$$\text{COV} - \text{R}(S_g, S_r) = \frac{1}{|S_r|} |\{C \in S_r | RMSD(C, \hat{C}) \leq \delta, \hat{C} \in S_g\}. \quad (12)$$

$$\text{MAT} - \text{R}(S_g, S_r) = \frac{1}{|S_r|} \sum_{C \in S_r} \min_{\hat{C} \in S_g} RMSD(C, \hat{C}). \quad (13)$$

where $S_g$ denotes the ensemble of generated conformations, while $S_r$ represents the ensemble of true conformations. $C$ and $\hat{C}$ represent individual conformations from the sets of true and generated conformations, respectively, with $\delta$ acting as the threshold, set at 1.25 Å. The COV metric evaluates the percentage of structures in one set that are encompassed by another set, where inclusion indicates that the RMSD between two conformations falls below a specified threshold $\delta$. Conversely, the MAT scores gauge the average RMSD between conformers in one set and their nearest counterparts in another set. Precision, as described in the provided equation, interchanges $S_g$ and $S_r$. Consequently, while Recall entails comparing each true conformation with all generated conformations, Precision involves comparing each generated conformation with all true conformations. Precision typically accentuates quality, while Recall is more concerned with diversity.

**Pocket-based molecular generation. Valid:** validity of generated 3D structure, calculates as the proportion of 3D structures that can be translated into canonical SMILES.

**IntDiv:** internal diversity[51], an assessment of the distinctiveness of molecules within a molecular set, calculated using Tanimoto distance based on ECFP4 fingerprints[52, 53].

**Smi:** similarity between two molecular sets, calculate same as IntDiv.

**Vina score:** the binding energy of ligands to protein pockets by using QVina2[55].

**High score:** the average ratio of generated molecules exceeding the original molecule within each pocket.

**MW:** Molecular weights.

**LogP:** the octanol-water partition coefficient, typically falls within the range of -0.4 to 5.6 for the majority of druglike compounds[86].

**Lipinski:** Lipinski's rule-of-five[58], which consists of the following criteria: the molecular weight of the compound is less than 500 daltons; the number of hydrogen bond acceptors in the compound's structure (including hydroxyl and amino groups) does not exceed 5; the number of hydrogen bond donors in the compound does not exceed 10; the logarithm of the compound's logP falls within the range of -2 to 5. the number of rotatable bonds in the compound does not exceed 10.

**QED:** quantitative estimation of drug-likeness[56], subsequent researchers have normalized the properties of Lipinski's rule-of-five into continuous values ranging from 0 to 1, where higher values indicate higher drug-likeness of molecules.

**SA score:** synthetic accessibility score[57], the SA score represents the synthesis accessibility of molecules and is designated on a scale of 1 to 10, based on chemical expertise. A higher value indicates greater difficulty in synthesis.

**Molecular property prediction.** During the evaluation, we employ greedy decoding for property prediction. Each method is run independently three times, and the average and standard deviation are reported. We utiliz the area under the receiver operating characteristic curve (ROC-AUC) [87] metric to evaluate the classification datasets. For the regression datasets, root mean square error (RMSE) is used to quantify the average difference between predicted values and actual values, which is often applied in regression analysis.

**Dataset**

**Pretraining.** The pretraining dataset is sourced from the geometric ensemble of molecules (GEOM) dataset, which includes conformers for 317,000 species, augmented with experimental data spanning biophysics, physiology, and physical chemistry domains[88]. These conformers are generated using sophisticated sampling methods coupled with semi-empirical density functional theory (DFT). Following this, data curation procedures are implemented to exclude molecules containing heavy metals, lacking torsions, or test molecules. Subsequently, each molecule underwent pre-training with a maximum of 30 conformers, yielding a final dataset containing 8,452,080 entries.

**Pocket-based molecular generation.** The dataset utilized for pocket-based generation is the same as existing work, which is an open-available dataset consisted of over 20 million of pose pairs from nearly 20,000 protein-ligand complexes from CrossDock2020[89]. Following the protocol outlined in previous studies[49, 67], we discarded all poses with an RMSD greater than 2Å, and additionally partitioned the dataset into training and testing sets based on a principle of sequence similarity less than 40%, ensuring a fairer evaluation of the generalizability to unknown pockets. Additionally, we excluded protein-ligand pairs which ligand lacked torsion angles from the dataset, resulted in slightly smaller training and testing sets compared to several models we mention subsequently.

The real-world targets' structure are download from RCSB PDB[90], and reference molecules corresponding to each targets are collected from ChemBL30 database[91]. Molecules with a $K_d$ or $K_i$ value less than 1,000 nM for a given target are considered active, counting into the reference sets. If the number of molecules meeting this criterion is low, molecules with an IC50 value less than 1,000 nM are also included in the reference sets. The collected molecules are deduplicated based on SMILES and molecules containing salts are removed.

**Molecular conformation generation**. We performed fine-tuning using datasets consistent with those utilized in earlier studies[33, 37, 39, 41]. For the test set, we employed a dataset akin to Tora3D[26]. Test set I contains 200 molecules, each with fewer than 100 conformations. Test set II comprises 1,000 randomly selected molecules with conformation counts distributed similarly to the entire dataset, spanning from 0 to 500.

**Molecular Property prediction**. We assembled a comprehensive collection of 12 datasets sourced from MolecularNet[92] and therapeutics data commons (TDC)[93], accompanied by comprehensive datasets descriptions provided in the **Supplementary**. Drawing upon MolecularNet's established status as a primary benchmark for molecular property prediction, our selection comprised six classification datasets and three regression datasets. Furthermore, within TDC, widely acclaimed as the premier public benchmark for ADMET analysis, we specifically identified three datasets characterized by relatively homogeneous data distributions. Each dataset underwent three random partitions following the 8:1:1 ratio for testing.

# Data availability

The datasets utilized in our study are as follows: The GEOM dataset is available at https://dataverse.harvard.edu/dataset.xhtml?persistentId=doi:10.7910/DVN/JNGTDF. For pocket-based molecular generation dataset is provided at https://drive.google.com/drive/folders/1CzwxmTpjbrt83z_wBzcQncq84OVDPurM.

The molecular conformation generation, the dataset can be accessed at https://github.com/zimeizhng/Tora3D. Lastly, the datasets for property prediction are available at https://moleculenet.org/datasets-1 and https://tdcommons.ai/single_pred_tasks/adme/.

## Code availability

The code used in the study is publicly available from the GitHub repository: https://github.com/jkwang93/Token-Mol.

## Acknowledgements

This work was financially supported by National Key Research and Development Program of China (2022YFF1203003), National Natural Science Foundation of China (22220102001, 82204279, 81973281, 82373791, 22303083), China Postdoctoral Science Foundation (2023M733128, 2023TQ0285), Postdoctoral Fellowship Program of CPSF (GZB20230657).

## Author Contributions

T.J.H., C.Y.H. and Y.K. designed the research study. J.K.W. developed the method and wrote the code. J.K.W., R.Q., M.Y.W., M.J.F., Y.Y.Z., G.Q.L., Q. S. performed the analysis. J.K.W., R.Q., M.Y.W., M.J.F., Y.Y.Z., Y.K., C.Y.H. and T.J.H. wrote the paper. All authors read and approved the manuscript.

## Competing interests

The authors declare that they have no competing interests.

# Supplementary

## 1. Attention Analysis of molecular prediction task

In the main manuscript, we validated Token-Mol's robust predictive capabilities in molecular property prediction tasks. While predictive accuracy is essential for scientific research, it is equally crucial for a model to validate scientific hypotheses. To this end, we focused on drug solubility and acute oral toxicity, areas extensively explored and significant for verifying model interpretability[1]. Using the ESOL and LD50 datasets, we illustrated how Token-Mol can extract relevant features from specific atoms or functional groups and assign corresponding attention weights to related tokens.

The ESOL dataset relates to molecular solubility, where various functional groups significantly influence solubility. Molecules containing polar functional groups, such as hydroxyl, carboxyl, and amino groups, generally exhibit higher solubility[2] due to their ability to form hydrogen bonds with water molecules. In contrast, hydrophobic functional groups, such as alkyl and halogen groups, tend to reduce solubility. **Fig. S6a** shows that in hydrophilic substructures, sp2 hybridized carbon, hydroxyl groups, and amine groups substantially impact the molecule's attention weights. For low-solubility molecules, the figure indicates that chlorophenyl and halogen atoms receive high attention weights. Moreover, a comparison in **Fig. S6a** reveals that, unlike high-solubility molecules where single functional groups often have significant influence, in drug design, the hydrophobicity of a molecule typically has a more holistic impact, with modifications of key hydrophobic groups affecting the overall structure's hydrophobic properties.

In drug toxicity research, we typically focus on toxicophores, or structural alerts (SAs)[3], defined as key substructures responsible for specific toxicities and commonly used to elucidate these relationships. For example, in the LD50 dataset, we identified the nitrosamide and phosphoric trimester toxicophores as reported by Wu et al.[4]. Nitrosamide is a frequently occurring SA in both mutagenic and non-mutagenic compounds[5], while phosphoric trimester is commonly found in compounds with acute oral toxicity[6]. We filtered the dataset for molecules containing nitrosamide and phosphoric trimester toxicophores and visualized the model's attention weights for these molecules. As shown in **Fig. S6b**, the model indeed assigns high attention weights to these toxicophores in all molecules containing the identified toxic fragments.

## 2. Datasets Details

The descriptions of various benchmarks dataset are listed in **Table S6**.

- BACE is a database that catalogues molecules with 2D structures and properties, focusing on inhibitors of human β-secretase 1. It provides qualitative (binary label) binding results for these inhibitors.
- The BBBP dataset comprises molecules with measured permeability properties, specifically assessing their ability to penetrate the blood-brain barrier.
- ClinTox is a dataset that encompasses both FDA-approved drugs and compounds that have been discontinued from clinical trials due to toxicity concerns.
- SIDER is a database that catalogs marketed drugs along with their associated adverse drug reactions (ADRs), categorized into 27 system organ classes.
- Tox21 assesses the toxicity of compounds across 12 different targets, spanning nuclear receptors and stress response pathways. It served as a public database for the 2014 Tox21 Data Challenge.
- ToxCast comprises toxicology data for thousands of molecules, offering multiple toxicity labels derived from high-throughput screening experiments conducted on a vast chemical library.
- The ESOL (Free Solvation Energy of Neutral Organic Molecules in Water) dataset is a standard dataset used in machine learning and cheminformatics research. It aims to predict the free solvation energy of organic molecules in water.
- The Free Solvation Database, FreeSolv (SAMPL), provides experimental and calculated hydration free energy of small molecules in water. The calculated values are derived from alchemical free energy calculations using molecular dynamics simulations.
- Lipophilicity measures the ability of a drug to dissolve in a lipid (e.g. fats, oils) environment. High lipophilicity often leads to high rate of metabolism, poor solubility, high turn-over, and low absorption.

- The human colon epithelial cancer cell line, Caco-2, is used as an in vitro model to simulate the human intestinal tissue. The experimental result on the rate of drug passing through the Caco-2 cells can approximate the rate at which the drug permeates through the human intestinal tissue.
- Aqeuous solubility measures a drug's ability to dissolve in water. Poor water solubility could lead to slow drug absorptions, inadequate bioavailablity and even induce toxicity. More than 40% of new chemical entities are not soluble.
- Acute toxicity LD50 measures the most conservative dose that can lead to lethal adverse effects. The higher the dose, the more lethal of a drug.

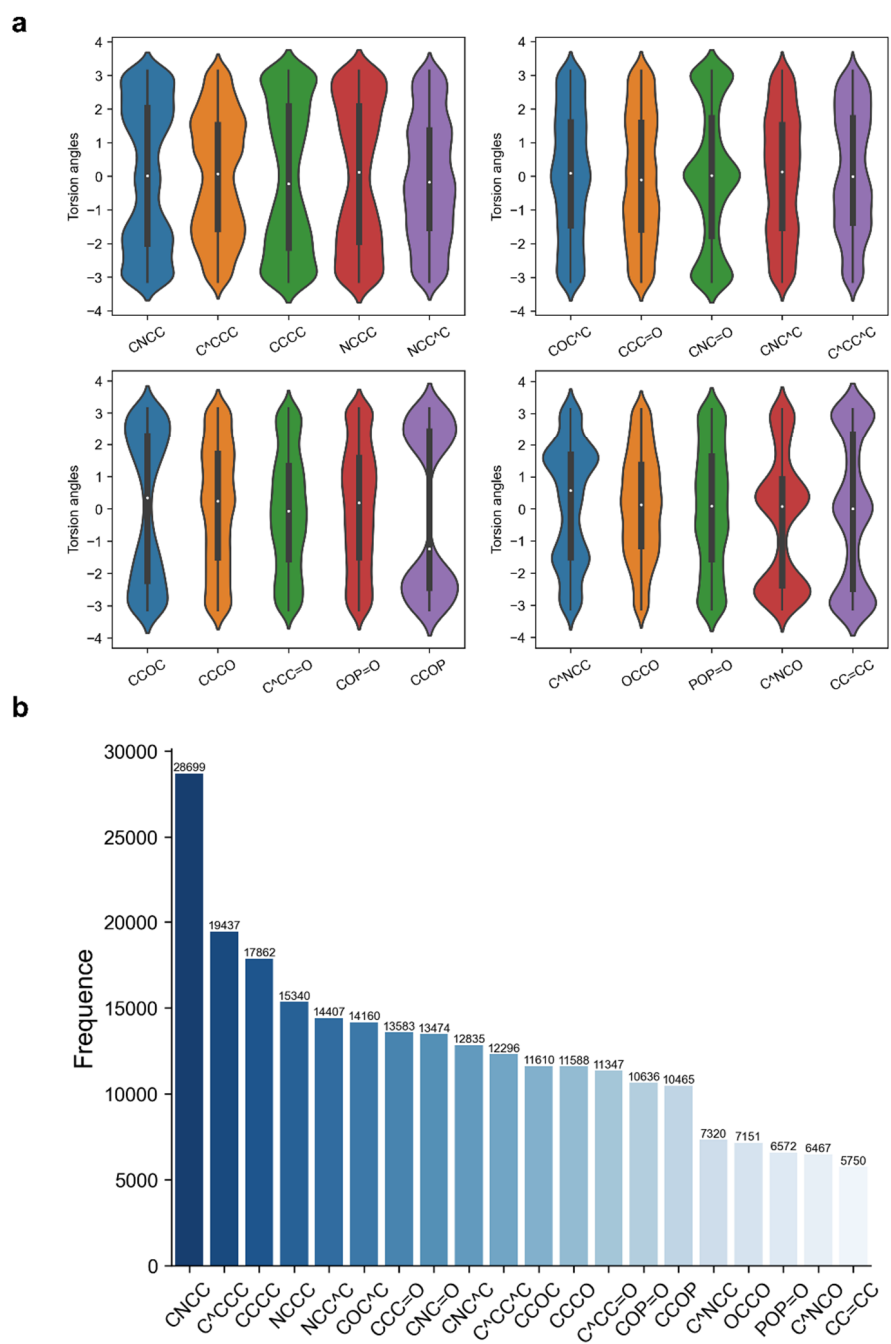

**Fig. S1 | The distributions of torsion angles in the training set.** (**a**) Values distributions (From -3.14 to 3.14) and (**b**) frequencies of the common torsion angles in the training set.

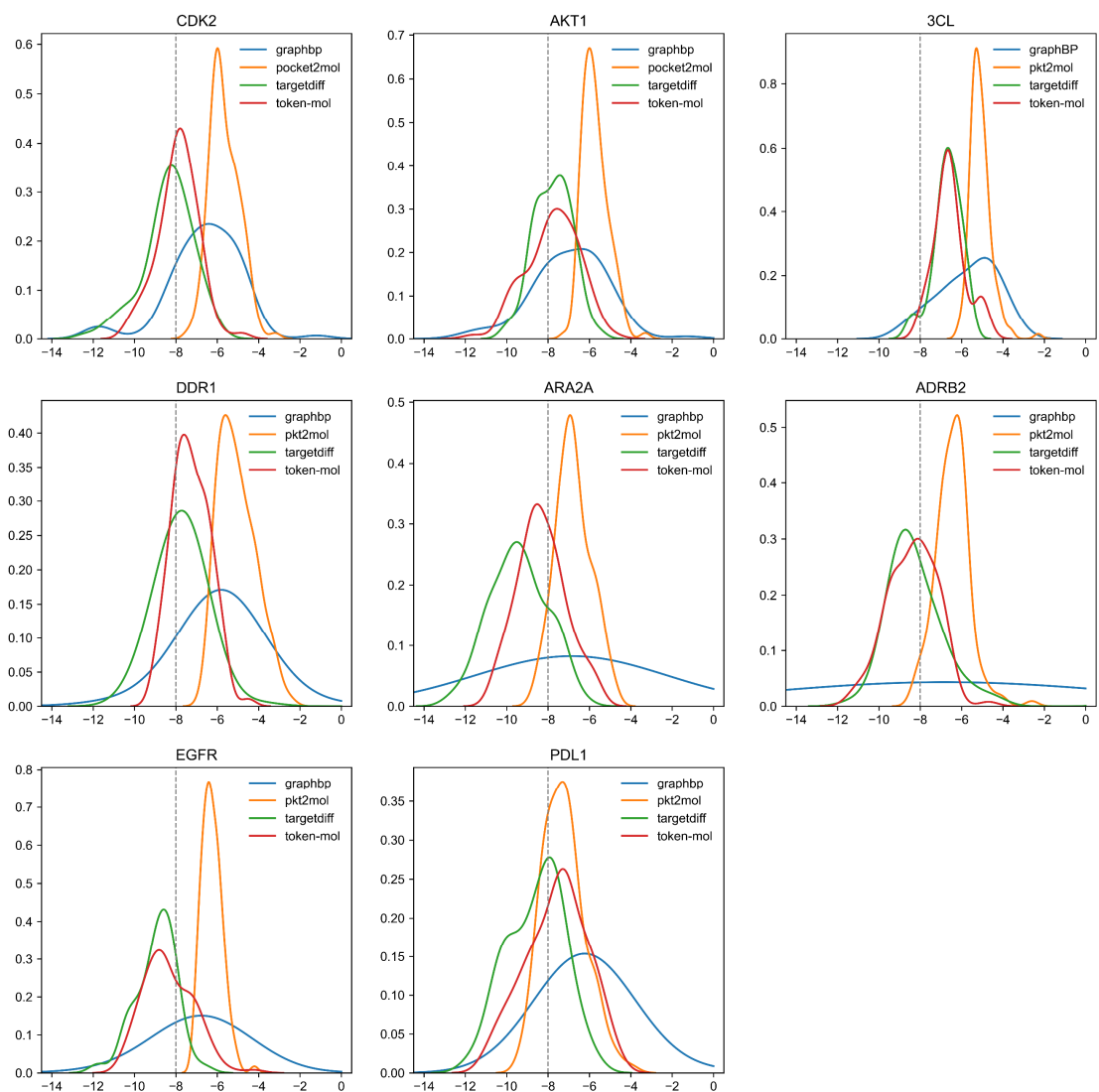

**Fig. S2 | Vina score distributions of Token-Mol and baseline models on the selected real-world targets.** The gray dashed line and the area to its left represent *'high affinity'* molecules, defined as molecules with a Vina score of less than or equal to -8.

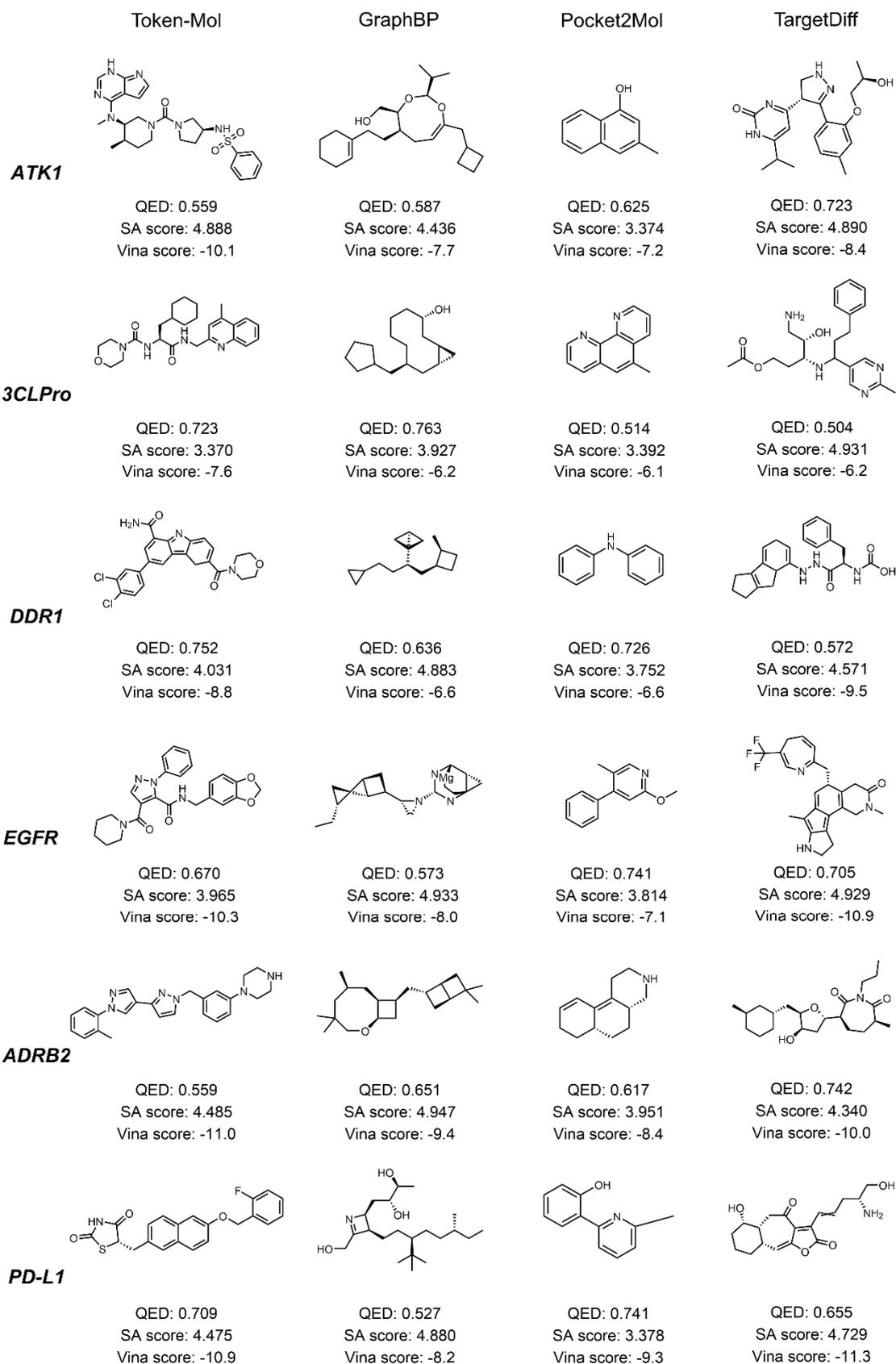

**Fig. S3 | The 'drug-like' molecules with the highest affinity and their properties on the selected real-world targets which are not represented in the main text.**

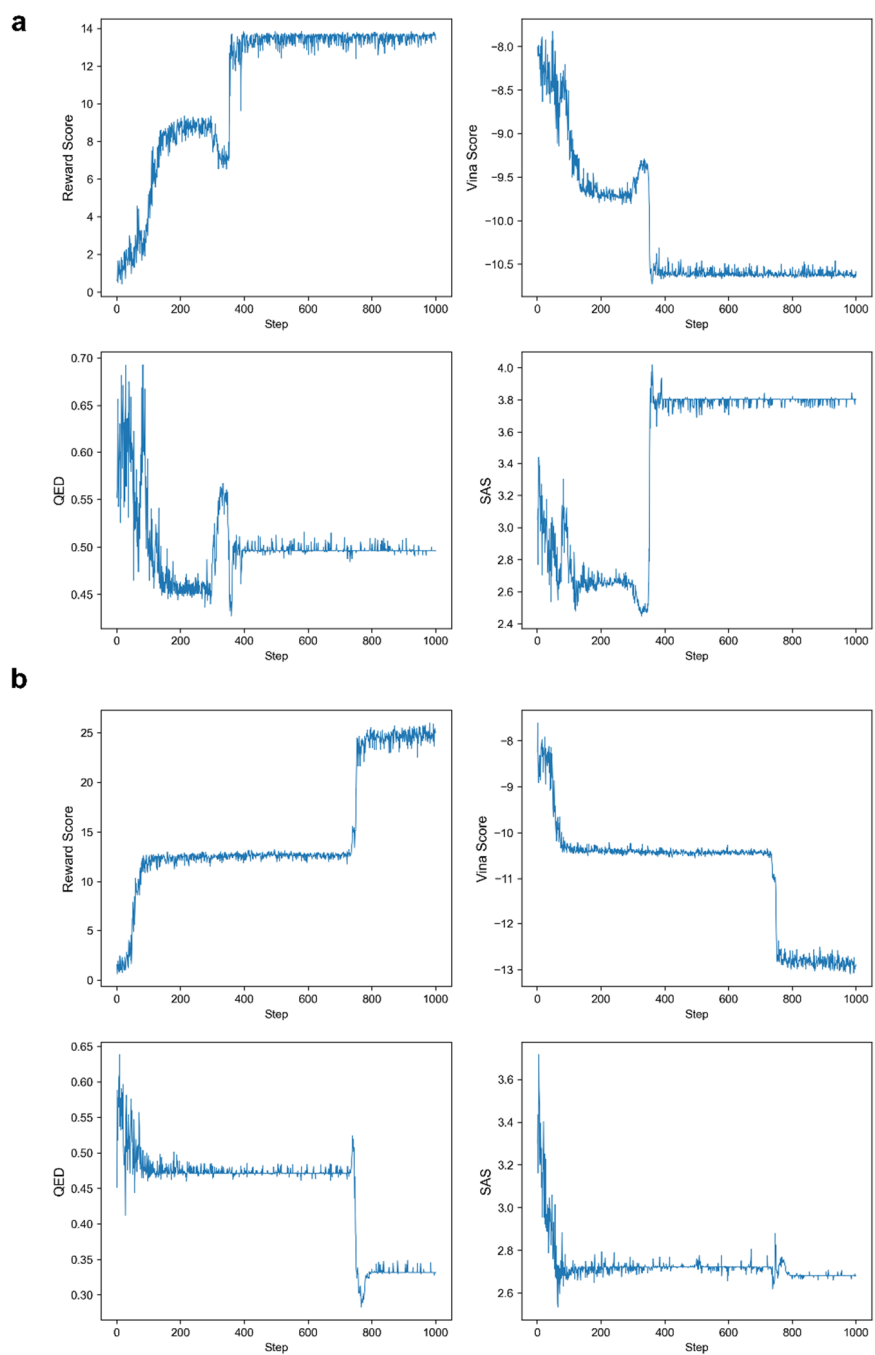

**Fig. S4 | Metrics during the reinforcement learning to optimized the binding affinity without any constraints in (a) CDK2 and (b) EGFR.**

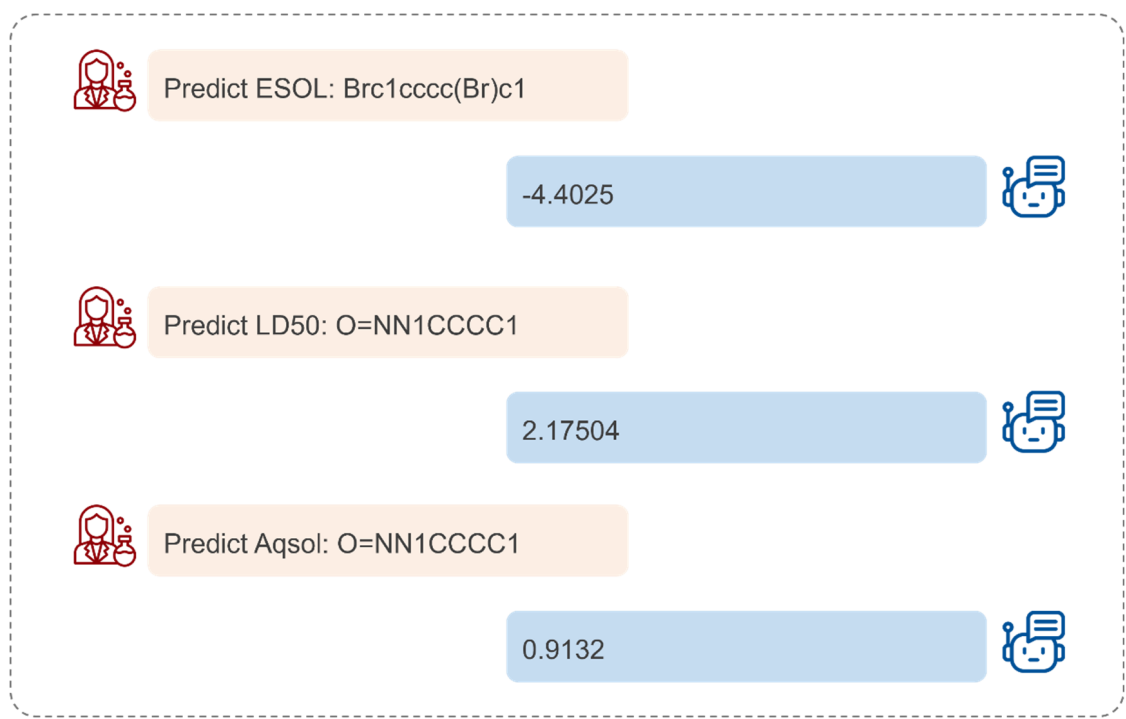

**Fig. S5 | The example of "Chat to Token-Mol".**

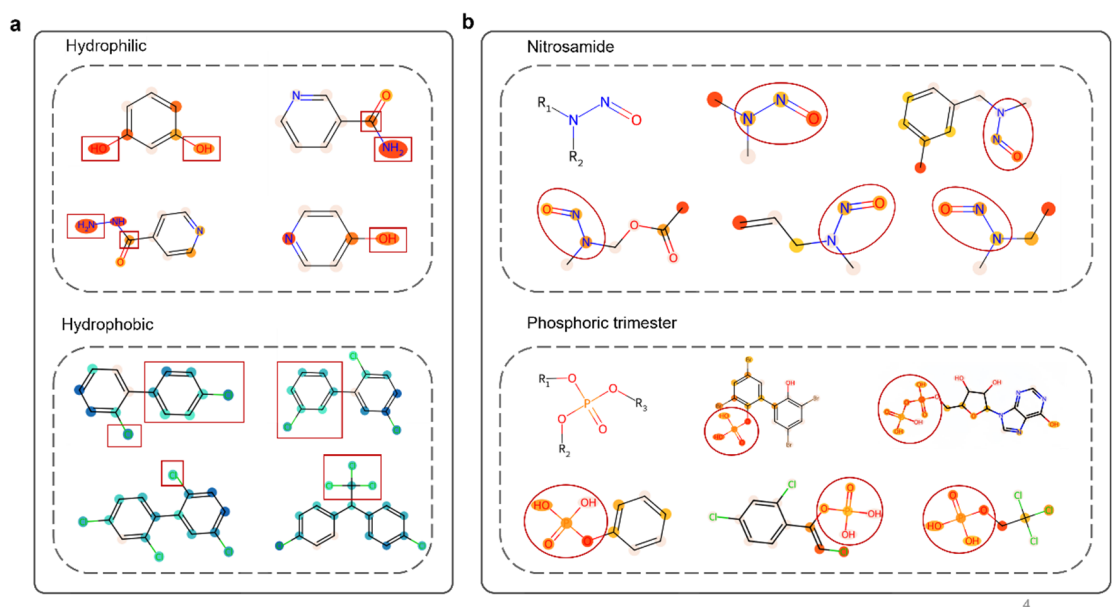

**Fig. S6 | Visualization of the attention weights for different molecules. (a)** The attribution visualization of aqueous solubility. The darker the color, the higher the weight. **(b)** Nitrosamide and phosphoric trimester of acute oral toxicity.

**Table S1 | Detailed data about the molecular properties of molecules generated in the test set by Token-Mol and other baseline models.**

| Metric | | Ori. | Token-Mol | GraphBP | Pocket2Mol | TargetDiff |
|---|---|---|---|---|---|---|
| Vina score ↓ | *Avg.* | 7.010 | -7.226 | -6.659 | **-7.729** | -7.236 |
| | *Med.* | -6.900 | -7.189 | -6.733 | -7.131 | **-7.251** |
| QED ↑ | *Avg.* | 0.492 | **0.568** | 0.511 | 0.536 | 0.493 |
| | *Med.* | 0.470 | **0.568** | 0.510 | 0.561 | 0.493 |
| SA score ↓ | *Avg.* | 4.214 | **4.166** | 5.271 | 5.846 | 4.858 |
| | *Med.* | 4.277 | **4.232** | 5.301 | 5.716 | 5.029 |
| Lipinski ↑ | *Avg.* | 4.247 | 4.666 | 4.749 | **4.782** | 4.586 |
| | *Med.* | 5.000 | 4.684 | 4.746 | **4.958** | 4.776 |
| MW | *Avg.* | 337.341 | 355.414 | 257.307 | 274.273 | 329.241 |
| | *Med.* | 326.013 | 366.857 | 257.772 | 242.405 | 340.350 |
| TPSA | *Avg.* | 114.292 | 87.320 | 39.834 | 72.246 | 94.945 |
| | *Med.* | 87.300 | 87.660 | 39.811 | 67.345 | 88.997 |

**Table S2 | The rate of high affinity molecules and the average QED values of generated molecules from different models on the selected real-world targets.**

|       | High Affinity (%) [#] | | | | QED | | | |
|-------|-----------|---------|-----------|------------|-----------|---------|-----------|------------|
|       | Token-Mol | GraphBP | Pocket2Mol | TargetDiff | Token-Mol | GraphBP | Pocket2Mol | TargetDiff |
| CDK2  | 33.97 | 15.29 | 0     | **58.60** | **0.630** | 0.464 | 0.578 | 0.472 |
| AKT1  | 36.78 | 22.35 | 0     | **39.13** | **0.592** | 0.481 | 0.578 | 0.472 |
| EGFR  | 63.96 | 15.66 | 0     | **85.90** | 0.537 | 0.433 | **0.614** | 0.282 |
| DDR1  | 18.52 | 6.82  | 0     | **38.46** | **0.589** | 0.430 | 0.585 | 0.331 |
| ARA2A | 61.82 | 21.18 | 7.34  | **80.00** | 0.560 | 0.447 | **0.628** | 0.618 |
| ADRB2 | 61.11 | 19.32 | 2.13  | **65.93** | 0.571 | 0.478 | **0.641** | 0.324 |
| 3CL   | 1.85  | **7.69** | 0  | 6.35  | 0.554 | 0.455 | **0.660** | 0.247 |
| PD-L1 | 36.11 | 4.34  | 23.01 | **58.90** | 0.553 | 0.457 | **0.620** | 0.391 |

[#] High affinity means Vina score lower than -8 (See also Supplementary Data Fig. S2).

**Table S3 | Detailed data for drug-like molecules with the highest affinity generated for the two selected targets from different models.**

| Target | Model | Vina Score | QED | SA Score | Lipinski | TPSA |
|---|---|---|---|---|---|---|
| CDK2 | Token-Mol | -9.4 | 0.746 | 4.434 | 5 | 92.26 |
| | Token-Mol (RL) | -10.1 | 0.689 | 3.830 | 5 | 68.10 |
| | GraphBP | -8.1 | 0.836 | 4.102 | 5 | 37.30 |
| | Pocket2Mol | -7.4 | 0.625 | 3.374 | 5 | 20.23 |
| | TargetDiff | -8.9 | 0.586 | 4.868 | 5 | 112.74 |
| ARA2A | Token-Mol | -9.5 | 0.541 | 4.306 | 5 | 49.69 |
| | Token-Mol (RL) | -10.3 | 0.701 | 2.567 | 5 | 58.10 |
| | GraphBP | -8.0 | 0.527 | 4.367 | 5 | 0 |
| | Pocket2Mol | -7.1 | 0.741 | 3.814 | 5 | 0 |
| | TargetDiff | -11.2 | 0.792 | 4.809 | 5 | 58.89 |

**Table S4 | The similarity between the generated molecules and reference molecules.**

|  | Scaffold Similarity↑ | | | | FCD↓ | | | |
|---|---|---|---|---|---|---|---|---|
|  | Token-Mol | GraphBP | Pocket2Mol | TargetDiff | Token-Mol | GraphBP | Pocket2Mol | TargetDiff |
| CDK2  | **0.121** | 0.026 | 0.099 | 0.101 | **42.863** | 52.530 | 47.603 | 46.167 |
| AKT1  | **0.101** | 0.039 | 0.083 | 0.097 | 51.058 | 51.327 | **48.625** | 53.346 |
| 3CL   | **0.103** | 0.050 | 0.083 | 0.086 | **37.741** | 46.988 | 44.441 | 40.010 |
| DDR1  | **0.127** | 0.034 | 0.051 | 0.111 | **47.912** | 56.974 | 55.331 | 49.085 |
| ARA2A | **0.111** | 0.033 | 0.096 | 0.102 | **53.982** | 64.379 | 61.755 | 54.312 |
| ADRB2 | **0.106** | 0.038 | 0.071 | 0.082 | 43.835 | 48.020 | 47.028 | **43.131** |
| EGFR  | **0.110** | 0.037 | 0.090 | 0.102 | 66.625 | 68.195 | **62.690** | 70.251 |
| PD-L1 | **0.106** | 0.029 | 0.072 | 0.094 | **47.734** | 50.706 | 49.660 | 48.366 |

Scaffold similarity is the similarity of the Bemis-Murcko scaffolds between reference molecules and generated molecules, calculated with ECFP4 fingerprints. FCD, Fréchet ChemNet Distance.

**Table S5 | Information about the reference molecules.**

| Target | Number | QED | Mean Vina Score |
|---|---|---|---|
| CDK2  | 402  | 0.482 | -8.589 |
| AKT1  | 156  | 0.461 | -7.612 |
| 3CL   | 448  | 0.438 | -6.966 |
| DDR1  | 261  | 0.401 | -8.010 |
| ARA2A | 3679 | 0.533 | -8.695 |
| ADRB2 | 457  | 0.405 | -8.977 |
| EGFR  | 421  | 0.482 | -8.247 |
| PD-L1 | 52   | 0.408 | 4.515  |

**Table S6 | Dataset details of molecular property perdition task.**

| Task Type | Metric | Dataset | Benchmark | Compounds | Tasks |
|---|---|---|---|---|---|
| Classification | ROC-AUC | BACE | MolecularNet | 1513 | 1 |
| | | BBBP | MolecularNet | 2039 | 1 |
| | | ClinTox | MolecularNet | 1478 | 2 |
| | | SIDER | MolecularNet | 1427 | 27 |
| | | Tox21 | MolecularNet | 7831 | 12 |
| | | ToxCast | MolecularNet | 8575 | 617 |
| Regression | RMSE | ESOL | MolecularNet | 1128 | 1 |
| | | FreeSolv | MolecularNet | 642 | 1 |
| | | Lipophilicity | MolecularNet | 4200 | 1 |
| | | Caco-2 | TDC | 906 | 1 |
| | | AqSolDB (Solubility) | TDC | 9982 | 1 |
| | | Acute Toxicity LD50 | TDC | 7385 | 1 |